\documentclass[11pt]{article} 
\usepackage[utf8]{inputenc}
\usepackage{multirow}
\usepackage{amsfonts}
\usepackage{amsmath}
\usepackage{amssymb}
\usepackage{amsthm,amsbsy,bm}
\usepackage{stackrel}

\usepackage{caption}
\usepackage[dvipsnames]{xcolor}
    \definecolor{darkgreen}{rgb}{0,0.5,0}
    \definecolor{darkblue}{rgb}{0,0,0.6}
    \definecolor{purple}{rgb}{0.4,.2,0.7}
\usepackage[margin = 2.5cm]{geometry}
\usepackage{graphicx}
\usepackage[hyperfootnotes = false, colorlinks = true, linkcolor = darkblue, citecolor = purple]{hyperref}
\usepackage{subcaption}
\usepackage{youngtab}

\newcommand{\be}{\begin{equation}}
\newcommand{\ee}{\end{equation}}
\newcommand{\bea}{\begin{eqnarray}}
\newcommand{\eea}{\end{eqnarray}}

\newcommand{\bc}{\begin{center}}
\newcommand{\ec}{\end{center}}

\newcommand{\ba}{\begin{eqnarray}}
\newcommand{\ea}{\end{eqnarray}}
\newcommand{\n}{\nonumber}
\newcommand{\nin} {\noindent}
\newcommand{\nn}{\nonumber}

	\newcommand{\bes}{\begin{equation} \begin{split} }	
	\newcommand{\ees}{\end{split} \end{equation} }

	\usepackage{physics}

\def\bps{\mathsf{TFD}}

\begin{document}

\thispagestyle{empty}
\begin{center}
    ~\vspace{5mm}

     
%
%
%
%
%
%
  {\LARGE \bf {Fermionic Matrix Models and  Bosonization  \\}}



   \vspace{0.5in}
     
   {\bf Matías N. Semp\'e and Guillermo A. Silva$^{1,2}$ }

    \vspace{0.5in}

   $^1$Instituto de F\'isica La Plata - CONICET 
   \\
   
   ~
   \\
   $^2$Departamento de F\'isica, Universidad Nacional de La Plata, \\
C.C. 67, 1900 - La Plata, Argentina

    \vspace{0.5in}

    \vspace{0.5in}
    

\end{center}

\begin{abstract}
We explore different limits of exactly solvable vector and matrix fermionic quantum mechanical models with quartic interactions at finite temperature. The models preserve a $U(1)\times SU(N)\times SU(L)$ symmetry at  the classical level and we analyze them through bosonization techniques introducing scalar (singlet) and matrix (non-singlet) bosonic fields. The bosonic path integral representations in the vector limits $(N,1)$ and $(1,L)$ are matched to fermionic Fock space Hamiltonians expressed in terms of  quadratic Casimirs and some additional terms involving the Cartan subalgebra, which makes explicit the symmetries preserved by scalar and matrix bosonizations at the quantum level. For the case of non-singlet bosonization we find  an equivalence between the vector model and the Polychronakos+Frahm spin model. Using this relation  we compute the free energy. Finally, we compute  the eigenvalue distribution  in the large $N,L$-limit with $ \alpha =  \frac{L}{N}$ fixed.  The model displays a third order phase transition as we vary the temperature which, in the  $\alpha\gg1$ limit,  can be  characterized  analytically.  We conclude  finding the critical curve in the parameter space  were the eigenvalue distribution transitions from single to double cut.
\end{abstract}

\vspace{1in}

\pagebreak

\setcounter{tocdepth}{3}
{\hypersetup{linkcolor=black}\tableofcontents}

  \section{Introduction} 

The study of matrix models has provided fruitful insights  within a vast number of physical problems, some examples   are \cite{Wig}: heavy nuclei, quantum chaos, gauge theories, string theory and 2d quantum gravity (see \cite{AM} for a recent review). 

In the present paper we will focus on random matrix theory with fermionic degrees of freedom  (for previous works along these lines see \cite{MZ},\cite{SemS},\cite{A15},\cite{AS}). Our main concern will be quantum mechanical  models with fermionic dof transforming in the bi-fundamental representation of $SU(M)\times SU(N)$ symmetry at finite temperature. An important motivation of our work is the discussion of the bosonization technique, in particular in relation to regularization ambiguities. As we will see,   non-single bosonization will preserve an Abelian subgroup of the classical symmetry group. On a broad context, our models, being finite dimensional, serve as nice playgrounds  to study   the proposal of a finite dimensional Hilbert space description for the static patch of a de Sitter universe  \cite{B00},\cite{W01}\cite{Li},\cite{BM},\cite{PV},\cite{DT}.  The study of fermionic tensor models has also recently gained attention due to two  reasons: on the one hand, a new large $N$ solvable limit was shown to exist for tensor models \cite{Gurau}; subsequently,  these large $N$ limits were further explored in \cite{kleba} in connection to the SYK model \cite{syk}-\cite{syk2}.

As is well known, since the original work of t' Hooft, significant simplifications occur when performing large $N$ limits. The importance of the  models to be discussed in the present work relies in their solvability; this allows us to perform several limits with precise analytic control. Previous work by one of the authors \cite{AS} established a relation between the present models and the Chern+Simmons matrix model. This  relation was  exploited in  \cite{T17}, allowing us to  solve the models analytically for arbitrary $N$ and $L$. As astonishing as this is, we still lack a deeper understanding of  important aspects as the large $N,L$-limits, the phases diagram and dualities with other physical models, etc. This is the second motivation for our work. We remark that the study of phase transitions in matrix models has a long history starting with the seminal work of Brezin+Itzykson+Parisi+Zuber  \cite{BIPZ} and Gross+Wadia+Witten \cite{GWW}  (see also \cite{shi},\cite{molinari},\cite{J90},\cite{Ake96} and the superb lecture notes \cite{M04}). For some recent studies of phase transitions in matrix models  see \cite{R20}.

The structure of the paper goes as follows: we start by analyzing the vector model from a group theory perspective. In section \ref{2} we bosonize the model performing the Hubbard+Stratonovich trick  and compute the partition function in closed form. We then proceed to show that the vector model spectrum  can be reproduced, in operator formalism, in terms of the quadratic Casimir of the $SU(N)$ symmetry with degeneracies  matching the dimensions of  $SU(N)$ irreps. In section \ref{3} we  study the matrix model, obtained from a non-singlet bosonization, for arbitrary $L$ and $N$. We review the results obtained in \cite{AS},\cite{T17} and  discuss in detail the  $L=1$ case. We will  find  that  non-singlet bosonization leads to a (vector model) Hamiltonian which can be expressed as the quadratic Casimir plus some additional terms belonging to the Cartan sub algebra. In this way we identify precisely the order ambiguities fixed by the bosonization. In section \ref{FPs} we further relate the Hamiltonian found in sect. \ref{L=1} to a known spin system (Frahm+Polychronakos (FP) model) computing and plotting the entropy for arbitrary $N$.  In section \ref{LNs}  we study several large $N$ limits of the models. As an outcome, the results of the previous section allows to give a matrix model description of the FP spin system. We also evaluate  the large $N,L$ limit keeping $\alpha=L/N$ constant, compute the eigenvalue distribution and find a third order phase transition . The transition curve in parameter space is evaluated numerically and plotted in section \ref{PTc}. We close with a summary and leave conventions and a review of the $\mathbb Z_2$-symmetric quartic model to the appendices.

\section {  Fermion vector model }

  \label{2}

\nin To setup the notation we start recalling that a QM fermionic partition function $Z(\beta)={\rm tr}(e^{-\beta H})$ is described in the path integral formalism as 
\be
Z(\beta)={\cal N}\int D\psi  D\bar\psi e^{-\oint[\bar\psi \dot\psi +H(\psi,\bar\psi)]}~\quad~\text{with AP fermions } ~\quad~\psi (\tau+\beta)=-\psi (\tau). 
\label{model}
\ee
Here $\psi$ denote complex classical Grassmann variables 
$$\{\psi^i,\psi^j\}=\{\bar\psi_i,\bar\psi_j\}=\{\psi^i,\bar\psi_j\}=0.$$ 
with  $\bar\psi_i=(\psi^i)^\dagger$ $(i=1,...,N)$.
We will consider the {\it classical}   Hamiltonian  
\be
 H=J\, (\bar\psi_i\psi^i)^2
\label{Hv}
\ee
 and discuss   bosonizations of the model using the  scalar  and matrix fields $\sigma$ and $A^i{}_{j}$. Eventually we will  express the thermodynamic quantities  rewriting the coupling as $J\propto 1/\tilde\gamma$. Dimensional analysis tells us that fermion fields are dimensionless, $[\psi]=0$, while  auxiliary fields $\sigma, A$ and the coupling constant have dimensions of energy, $[\sigma]=[A]=[J]=E~$, $[\tilde\gamma]=E^{-1}$. As a consequence, there is a single dimensionless coupling on which the partition function can depend, i.e. $\beta/\tilde\gamma$.   Dimensionless  bosonic fields $\lambda, M$ will show up at intermediate steps, they are defined  as   $\lambda\equiv \beta\sigma,~M\equiv\beta A$.  A high temperature regime is equivalent to large $\tilde\gamma$ coupling. Eventually we will compare the path integral results with the Fock space perspective. Fermionic creation (annihilation) operators will be denoted in upper case $\bar\Psi\,(\Psi)$.

\subsection{Bosonization with a singlet} 
\label{bs}

We choose the coupling $J$ to be negative and rewrite it as  $ J=-\frac1{4N\tilde\gamma}<0$ with $\tilde\gamma>0$.  The power of $N$ in the denominator will guarantee a sensible large $N$-limit.  The standard Hubbard+Stratonovich trick
\be
\int D\sigma e^{ -\oint  (N{\tilde\gamma}\sigma^2+\sigma  \bar\psi_i\psi^i)}=e^{ \oint \frac1{4N{\tilde\gamma}} \bar\psi_i\psi^i\bar\psi_j\psi^j}
\label{bsn}
\ee 
inserted in the partition function \eqref{Hv} gives, after integrating out the fermions \cite{AS},
\begin{align}
  Z_N\left( \beta/ {\tilde\gamma}\right)& ={\cal N}\int D\sigma\,\det\phantom{{}}^N(\partial_\tau+\sigma)e^{-N\tilde\gamma\oint\sigma^2} \n\\
 &={\cal N}\int d\sigma_0\,\cosh^N\frac{\sigma_0}2\,e^{-N\frac{\tilde\gamma}\beta\sigma_0^2}\n  \\
 &=\sum_{k=0}^N  {N\choose k}\,  e^{-\beta E_k}
 \label{ptf}
\end{align}
From this expression we read the system energies
\be
 E_k=-\frac{(N-2k)^2}{16N\tilde\gamma },~~~~~~~~~~~~k=0,...[N/2]
 \label{ener}
 \ee
and degeneracies
\begin{align}
N\text{ odd}:~~~~d_k=
 2{N\choose k},~~\forall k~~~~~~~~~~N\text{ even}:~~~d_k&=
 2{N\choose k},~~~\forall k\ne N/2, ~~~~  d_{N/2}= {N\choose k} 
 \label{spec}
 \end{align}
The factor of 2 in the degeneracies arises because \eqref{ptf} enjoys  a `{\it reflection}' $k\to N-k$ symmetry,  reducing the number of terms in the sum  by half.

\vspace{2mm}

A few comments are in order: 

\nin(i)  The normalization constant $\cal N$ in \eqref{ptf} was adjusted so that in the infinite temperature limit we get  the dimension of the Hilbert space,
$$Z_N(0)=2^N~\leadsto ~{\cal N}=\frac{2^N}{\int d\sigma_0 \,e^{-N\frac{\tilde\gamma}\beta\sigma_0^2}}. $$
(ii) Gaussian fermions Yukawa coupled to  $\sigma$ have a local reparametrization invariance
\be
\tau\to f(\tau),~~~~\sigma(\tau)\to\frac1{f'(\tau)}\sigma(\tau),~~~~\psi(\tau)\to\psi(\tau).
\label{gs}
\ee
This (gauge) invariance erases all $\tau$-dependence from $\sigma$, and fermion dynamics reduces to that of Matsubara zero mode $\sigma_0=\oint d\tau\,\sigma(\tau)$. This is the only gauge invariant object that we can construct out of $\sigma(\tau)$.\\
(iii) The cosh in the second line of \eqref{ptf} arises from the determinant  computed on anti-periodic functions. We have chosen to preserve a $\mathbb Z_2$  symmetry, i.e. $\sigma\to-\sigma$ in \eqref{ptf} (cf. App.D in \cite{SK} for further discussions)\footnote{See \cite{BT} for related recent work.}.\\
(iv) The $N$ factor in $J$ is tuned to have a competition, at large $N$, between the Gaussian and $\cosh$ in the second line of \eqref{ptf}.\\
(v) The case $J>0$  (${\tilde\gamma}<0$) can be obtained by integrating $\sigma$ over the imaginary axis $\sigma\to i\sigma$.

\vspace{2mm}

\nin Some particular cases  are
\begin{eqnarray}
&Z_{2}=2e^{\frac\beta{\tilde\gamma}\frac18}+2&\to~~ 2^2 \n \\
&Z_{3 }=2e^{\frac\beta{\tilde\gamma}\frac3{16}}+6e^{\frac\beta{\tilde\gamma} \frac1{48}} &\to~~ 2^3 \n\\
&Z_{4}= 2e^{\frac\beta{\tilde\gamma}\frac{1}{4}}+8e^{\frac\beta{\tilde\gamma}\frac{1}{16}}+6&\to~~ 2^4 \n \\
&Z_{5}= 2e^{\frac\beta{\tilde\gamma}\frac{5}{16}}+10e^{\frac\beta{\tilde\gamma}\frac{9}{80}}+20e^{\frac\beta{\tilde\gamma}\frac{1}{80}}  &\to~~  2^5 \n \\
&Z_{6}= 2e^{\frac\beta{\tilde\gamma}\frac{3}{8}} +12e^{\frac\beta{\tilde\gamma}\frac{1}{6}}+30e^{\frac\beta{\tilde\gamma}\frac{1}{24}}+20&\to  ~~2^6 \n \\
&Z_{7}=\hspace{-2mm}\underbrace{\, 2e^{\frac\beta{\tilde\gamma}\frac{7}{16}}}_{\text{\sf ground state}}\hspace{-3mm}+14e^{\frac\beta{\tilde\gamma}\frac{25}{112}}+42e^{\frac\beta{\tilde\gamma}\frac{9}{112}}+\hspace{-6mm}\underbrace{\,70e^{\frac\beta{\tilde\gamma}\frac{1}{112}}}_{\text{\sf highest energy state}}&\to \underbrace{2^7}_{\beta\to0}\n 
\end{eqnarray}
{\sf Fock space perspective}: The energy spectrum  \eqref{ener} can be obtained from the {\it quantum} Hamiltonian \cite{AS}
\begin{align}
  \hat H&=J\left( \bar\Psi_i\Psi^i\bar\Psi_j\Psi^j-\bar\Psi_i\Psi^iN+\frac {N^2}4\right)\n\\
  &=-\frac1{4{\tilde\gamma} N}\left( \bar\Psi_i\Psi^i\bar\Psi_j\Psi^j-\bar\Psi_i\Psi^iN+\frac {N^2}4\right),~~~~~~~~~~i,j=1,...,N
\label{H}
\end{align}
acting on the Fock space built out of the Grassmann operators  
$$\{\Psi^i,\bar\Psi_j\}=\delta^i_j. $$  

It is important to realize that in the transition to the quantum theory, the coefficients of the last two terms in \eqref{H} are ambiguous. As we show below, the bosonization process fixes them so that the quantum Hamiltonian $\hat H$  preserves the classical $U(N)=U(1)\times SU(N)$ symmetry enjoyed by \eqref{Hv}  
\begin{align}
U(1):&~~~~\psi^i\to e^{i\alpha}\psi^i\nn\\
SU(N):&~~~~\psi^i\to U^ i{}_ j\, \psi^j~~~~~~\text{with }~~~U\in SU(N)\label{su}
\end{align}
Modulo a constant shift, $\hat H$ can be obtained by replacing the {\it classical\,} Grassmann variables $\psi^i$ by operators  $\Psi^i$ in the classically equivalent Hamiltonian
\begin{align}
H=-J\,\bar\psi_i\psi^i\psi^j\bar\psi_j~\quad\mapsto\quad~\hat H'&=-J\,\bar\Psi_i\Psi^i\Psi^j\bar\Psi_j\nn\\
&=J\left(\bar\Psi_i\Psi^i\bar\Psi_j\Psi^j-N\bar\Psi_i\Psi^i\right).
\end{align}

\subsection{$U(N)$ symmetry and Fock space decomposition}
  
The Fock space of the $N$-fermion system consists of $2^N$ states build out from creation and annihilation operators satisfying 
\be
\{\Psi^i,\bar\Psi_j\}=\delta^ i_j ,~~i,j=1,...N.
\label{AR}
\ee
As customary  we define the vacuum state $|0\rangle$  as 
$$\Psi^i|0\rangle=0,~\qquad\forall i$$ 
and construct totally antisymmetric states
\be
|i_1i_2..i_{k}\rangle\equiv  \bar\Psi_{i_1}...\bar\Psi_{i_k}|0\rangle~~~\Rightarrow~~~ |i_1i_2..i_n....i_m...i_{k}\rangle=- |i_1i_2..i_m....i_n...i_{k}\rangle~~~\forall~m,n
\label{fst}
\ee
It is straightforward to see that there are $ {N\choose k}$ distinct states at fixed  $k$ arising from index combinatorics. 

\vspace{2mm}
\nin. $U(1)$: we can classify the Fock states \eqref{fst} by its $U(1)$ charge,
\be
U(1)\text{ charge}:~~~~{\sf Q}=\sum_i\bar\Psi_i\Psi^i~.
\label{u1}
\ee
This charge counts the number of fermions acting on the vacuum. Moreover, the Hamiltonian \eqref{H} can be written in terms of the $U(1)$ charge as
\begin{align}
\hat H&=-\frac1{4{\tilde\gamma} N}\left( {\sf Q}^2-{\sf Q}N+\frac {N^2}4\right)\nn\\
&=-\frac1{4{\tilde\gamma} N}\left( {\sf Q} - N/2 \right)^2.
\label{u1H}
\end{align}
Since $[\hat H,{\sf Q}]=0$  energy eigenstates  have a fixed number of fermions acting on the vacuum.  We now turn to the degeneracy issue.

\vspace{2mm}

\nin . $SU(N)$:    complex fermions $\Psi^i$ transform under the fundamental representation of $SU(N)$ (cf. \eqref{su}). Charges
\be
Q_\alpha=\sum_{i,j}\bar\Psi_i(T_\alpha)^i{}_{ j}\Psi^j,
\label{Qsu}
\ee
with $\bm T_\alpha$ the $SU(N)$ generators in the fundamental representation (see appendix for conventions), implement, at the quantum level, the $SU(N)$ transformations \eqref{su} as\footnote{The anti-commutation relations \eqref{AR} imply that charges \eqref{Qsu} close the $SU(N)$ algebra are the following: $[Q_\alpha,Q_\beta]= if^\gamma{}_{\alpha\beta}\,Q_\gamma $. }
$$\delta \Psi= -i[Q_\alpha,\Psi].$$
The $SU(N)$ action preserves the number of fermions in the state, {\it i.e.}, 
$$[{\sf Q},Q_\alpha]=0.$$
Thus, in view of \eqref{u1H} we conclude that
$$[\hat H,Q_\alpha]=0$$ 
Barring additional symmetries we expect energy levels degeneracies to coincide with the dimensions of $SU(N)$ irreps. 

\vspace{2mm}

\nin . {\sf Hamiltonian and Casimirs}:  Computing the quadratic $SU(N)$ Casimir \eqref{casiC} for the charges \eqref{Qsu} we find, using \eqref{fi},
\begin{align}
C^{(2)}=& Q_\alpha Q^\alpha =\bar\Psi_i(T_\alpha)^i{}_{ j}\Psi^j\, \bar\Psi_k(T^\alpha)^k{}_{ l}\Psi^l =-\frac{ N+1}{ 2N} \left( (\bar\Psi_k \Psi^k )^2-N\bar\Psi_i \Psi^i  \right)=-\frac{ N+1}{ 2N} \left( {\sf Q}^2-N\, {\sf Q} \right)
\label{SUNU1}
\end{align}
The final result is
 \begin{align}
\hat H&= -\frac1{4{\tilde\gamma} N}\left( {\sf Q}^2-{\sf Q}\,N+\frac {N^2}4\right)~~~~~~~~~\text{in terms of $U(1)$ Casimir }{\sf Q}\nn\\
 &=\frac1{2{\tilde\gamma} (N+1)} C ^{(2)}-\frac {N }{16{\tilde\gamma}} ~~~~~~~~~~~~~\text{in terms of $SU(N)$ Casimir }C ^{(2)}
 \label{casiH} 
\end{align}
Fermionic states at level $k$ comprise a completely antisymmetric irrep consisting in a single column with $k$ boxes, hence, using \eqref{genC},
$$C^{(2)}_{A_k}=\frac12\left(kN+k-k^2-\frac{k^2}N\right)=-\frac{N+1}{2N}\left(k^2 -kN \right)$$
Notice $A_k$ and $A_{N-k}$ have the same Casimir.  This analysis  shows that the Hilbert space can be decomposed in terms of $U(1)$ eigenstates, each of which is also a completely antisymmetric $SU(N)$ irrep
 \begin{align}
{\cal H}&=\bigoplus _{k=0}^N{\cal H}_{A_k}=1+{\tiny\yng(1)}+\begin{array}{c} {\tiny\yng(1,1)}\end{array}  +...+
\begin{array}{c}\tiny\yng(1,1)\\ \vdots\\ \tiny\yng(1)\end{array}
+1
\end{align}
and each subspace has degeneracy $d_n= {N\choose k}$.
 
 Notice that among the set of general Hamiltonians preserving $U(1)\times SU(N)$ 
$$H'=a \,C ^{(2)}+ b \,{\sf Q}^2+c {\sf Q}, $$
the regularization leading to \eqref{casiH} is singled out as being invariant under ${\sf Q}\mapsto (N-{\sf Q})$.

\subsection{Thermodynamics}

The thermodynamics of the  model was previously studied in \cite{AS}. The model displays in the large $N$-limit a confining/deconfining first order phase transition at $\beta_c/\tilde\gamma=8$ with entropy $S\sim {\cal O}(1)$ at low temperature and $S\sim {\cal O}(N)$ at high temperature. The order parameter for the phase transition is the   gauge invariant quantity  $\sigma_0$.

\begin{figure}[tbp]
\begin{center}
  \includegraphics[width=3.3in]{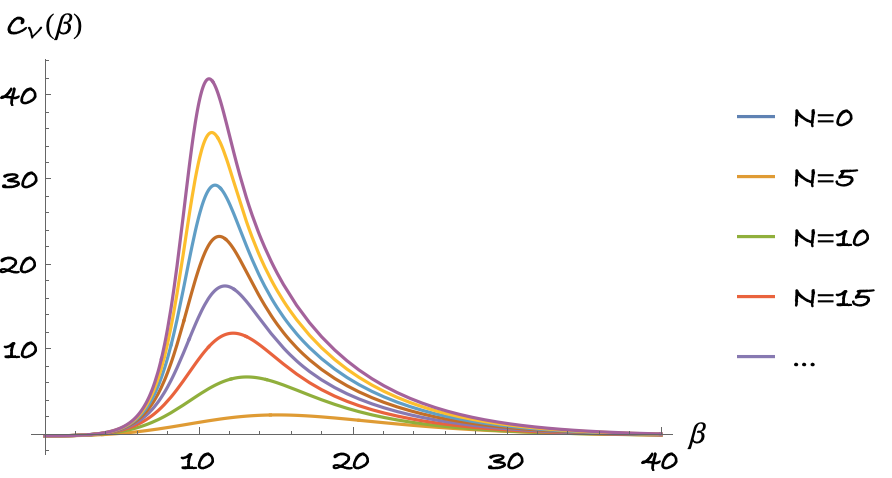}  
\end{center}
\caption{{\sf  Specific heat as a function of temperature}: $C_v$ displays a peak approaching $\beta_c= 8\gamma$ as $N$ grows (we set $\gamma=1$). The divergence in $C_v$ as $N\to\infty$ manifests a Hagedorn temperature for the model.} 
\label{cv}
\end{figure}

In the formal large $N$ limit, combining \eqref{ener} and \eqref{spec}, one finds the density of states as function of energy behaves as \cite{AS}
$$d(E)\sim e^{8\gamma E}.$$
This brings  a $1/(T-T_c)^2$ divergence in the specific heat $C_v=T\frac{\partial S}{\partial T}=-T\frac{\partial^2 F}{\partial T^2}$ which results in a limiting (Hagedorn) temperature for the model (see \cite{GKPP} for similar features in recent work).  The divergence in $C_v$ appears as a peak at finite $N$  as we approach $\beta_c= 8\gamma$ (see fig. \ref{cv})\footnote{We thank an anonymous referee for pointing out the Hagedorn behavior of our model and reference \cite{GKPP}.}.   
 
\section{Fermion matrix model}
\label{3}

We now turn to the fermion matrix model, built out of $N\times L$ complex Grassmann variables 
$$\{\psi^{iA},\bar\psi_ {Ai}\}=0$$ 
with $i = 1, . . . , N$ and $A = 1,...,L$ and $(\psi^{iA})^\dagger=\bar\psi_ {Ai}$. The  $A,B...$-indices   end up being spectators but provide an additional parameter, $L$, to play with.
 
The {\it classical} Hamiltonian  takes the form
\begin{align}
H&=-\frac1{4\tilde \gamma L}\left( \bar\psi_{Bi}\psi^{iA}\bar\psi_{Aj}\psi^{jB} \right).
\label{MF}
\end{align}
The $L$ factor in the coupling will give  rise to non-trivial thermal properties in the large $N,L$ limits. The model of interest to be discussed below corresponds to the analytic continuation $\tilde\gamma \to   -\tilde\gamma$.

\subsection{Bosonization with a matrix}

The model \eqref{MF} is bosonized by  writing the quartic term \eqref{MF} in terms of an auxiliary Hermitian matrix $A^i{}_j $ as
\be
\int DA\, e^{ -\oint  [L  \tilde\gamma\, \mathrm{tr}(A^2)-i \,\mathrm{tr}(A\psi  \bar\psi)]}=e^{- \oint \frac1{4L{\tilde \gamma}} \mathrm{tr}(\psi\bar\psi \psi \bar\psi)}.
\label{bsn2}
\ee 
Integrating out fermions one obtains
$$Z_{N\times L}=\int DA\,  \det\phantom{{}}^L(\partial_\tau+iA )\,e^{-L\tilde \gamma\oint  {\sf tr}A^2}.$$
The determinant is immediate to compute writing $A(\tau)=U(\tau)\cdot\bm\Lambda\cdot U^\dagger(\tau)-iU(\tau)\partial_\tau U^\dagger(\tau)$ with $\bm\Lambda={\sf diag}(\lambda_1,...\lambda_N)$ $\tau$-independent and  $U\in SU(N)$. The Jacobian for the change of variables $A\to(\lambda_i, U)$ can be found in \cite{AMV}. Performing an analytic continuation in the coupling $\tilde\gamma \to \gamma\equiv -\tilde\gamma$  in standard fashion, rotating the $\lambda$-contour to the imaginary axis \cite{Mar},\cite{W10},  the partition function finally reads \cite{AS}
\begin{equation}
Z_{N\times L}(\beta/\gamma)={\cal N}_{\beta,\gamma} \int \prod_{i=1}^N d\lambda_i \prod_{j>i} \sinh^2 \left(\frac{\lambda_i-\lambda_j} 2  \right)  \cosh^L \left(\frac{\lambda_i} 2 \right) e^{-\frac{L\gamma}\beta \lambda_i^2} .
\label{partition}
\end{equation}
The normalization ${\cal N}_{\beta,\gamma} $ is chosen to be
\be
{\cal N}_{\beta,\gamma}=\frac{2^{N\times L}}{\int \prod_{i=1}^N d\lambda_i \prod_{j>i} \sinh^2 \left(\frac{\lambda_i-\lambda_j} 2  \right)  e^{-\frac{L\gamma}\beta \lambda_i^2} },
\label{norm}
\ee
which in the high temperature limit gives\footnote{Notice there is a typo in (4.13) in \cite{AS}.}  $Z_{N\times L}[0]=2^{N\times L}={\sf dim} ({\cal H})$.

The  solution of the model was found in \cite{T17},  finding its  relation   to the Chern+Simons (CS) matrix integral. Performing the change of variables   $\log y_i =\lambda_i+\frac{\beta}{4\gamma}+\frac{N\beta}{2L\gamma}$ one can rewrite the partition function as
\be
Z_{N\times L}(\beta/\gamma)=\frac{e^{-\frac\beta\gamma( \frac{N^2}2+\frac{3N\,L}{16})}}{\int \prod_{i=1}^N {dy_i}  \prod_{j>i} (y_i -y_j )^2   e^{-L\frac\gamma\beta  \log^2 y_i  } } \int \prod_{i=1}^N  {d y_i} \prod_{j>i}    ( {y_i-y_j}   )^2  \left(y_i -a\right)^L e^{-\frac{L\gamma}\beta \log^2 y_i   }
\label{tie}
\ee
where $ a\equiv -e^{\frac{\beta}{4\gamma}+\frac{N\beta}{2L\gamma}} $. 

Through this reasoning \cite{T17} found that the partition function of the fermion matrix model coincides with the expectation value of the $L$-th power of the characteristic polynomial of an $N\times N$ Hermitian matrix $M$ in the Stieltjes+Wigert (SW) ensemble
\be
Z_{N\times L}=e^{-\frac\beta\gamma( \frac{N^2}2+\frac{3N\,L}{16})}\langle\,\big(\det (M- a\mathbb I)\, \big)^L\,\rangle_{SW}
\label{brezi}
\ee
Here, 
$$\langle F(M)\rangle_V\equiv\frac1{\cal Z}\int DM\, F(M)e^{-V(M)}~~~\quad~\text{and}~~~~~~~~~{\cal Z}=\int DM  e^{-V(M)}.$$
with SW  potential given by
$$V_{SW}(M)=\frac1 g\text{ \sf tr}\left[\big(\log M\big)^2\right]~~~~~\quad\text{ with }~~~~~~\frac 1g=\frac{L\gamma}\beta.$$

The expectation value of products of characteristic polynomials for Hermitian ensembles have been computed in \cite{BH} in terms of orthogonal monic polynomials. Applying the formulas to the present case (see App.\ref{poly}), the result is \cite{T17}
\be
 Z_{N\times L}=e^{-\frac\beta\gamma( \frac{N^2}2+\frac{3N\,L}{16})}\frac{(-)^{L(L-1)/2}(-)^{N\,L}}{\prod_{i=0}^{L-1}(i!)}\det\left|
\begin{array}{cccc}
p_N( a)&p_{N+1}( a)&...&p_{N+L-1}( a)\\
p'_N( a)&p'_{N+1}( a)&...&p'_{N+L-1}( a)\\
\vdots\\
p^{(L-1)}_N( a)&p^{(L-1)}_{N+1}( a)&...&p^{(L-1)}_{N+L-1}( a)
\end{array}\right|
\label{BHf}
\ee
where
$$a=-e^{\frac{\beta}{4\gamma}+\frac{N\beta}{2L\gamma}}. $$
Here $p_n(x)$  are orthogonal monic polynomials for the SW measure $d\mu=e^{-\alpha\log^2x}dx$ (see App.\ref{poly}). Their explicit expression is 
$$p_N(x)=(-)^Nq^{-N^2-N/2}\sum_{j=0}^N\binom{N}{j}_q(-)^jq^{j^2+j/2}x^j~~~~~~~~\text{with}~~~~q=e^{-g/2 }=e^{-\frac\beta{2L\gamma}}$$
The $q$-deformed binomial coefficient $\binom{N}{k}_q$ is defined as 
\begin{align}
\binom{N}{k}_q\equiv\frac{\prod_{r=1}^N (1-q^r)}{\prod_{r=1}^k (1-q^r)\prod_{r=1}^{N-k} (1-q^r)}=\prod_{r=1}^k\frac{ 1-q^{N-r+1 }}{  1-q^r }~\stackrel[q\to1]{}{\longrightarrow}~\binom{N}{k},
\end{align}
it is a polynomial in $q$.

\subsection{$L=1$  fermion  model}

\label{L=1}

The partition function \eqref{BHf} for $L=1$ reduces to a single polynomial and corresponds to the bosonization of the original model \eqref{Hv}, albeit with opposite sign $\tilde\gamma\to\gamma=-\tilde\gamma$ and  with a matrix $A^i{}_j$ instead of a singlet $\sigma$.   Since, $q=e^{-\frac\beta{2\gamma}}$ we then have $ a= -q^{-(N+1/2)}$. The partition function in the $L=1$ case takes the concise form
\begin{align}
 {Z}_{ N\times 1}&=q^{N^2+\frac38N}(-)^N p_N\left(-q^{-(N+1/2)}\right) \nn\\
&= q^{-N/8}\sum_{j=0}^N\binom{N}{j}_q q^{j^2-jN}  \nn\\
&= q^{-N^2/4-N/8}\sum_{j=0}^N\binom{N}{j}_q q^{(j-N/2)^2}  
\label{Z1T}
\end{align}
The exponent inside the sum is  invariant under reflection $j\mapsto (N-j)$ and so are the $q$-binomials. This implies again that the number of terms in the sum effectively reduces by half.  As recognized in \cite{T17}, since the number of distinct energies in the partition function is polynomial $\sim N^2$ while the number of states is exponential  $\sim 2^N$, we expect an exponential degeneracy growth.

Explicit examples are 
\begin{eqnarray}
&Z_{2\times1}=e^{\frac\beta\gamma\frac58}+3e^{\frac\beta\gamma\frac18}&\to~~ 2^2 \nn\\
&Z_{3\times1}=2e^{\frac\beta\gamma\frac{19}{16}}+2e^{\frac\beta\gamma\frac{11}{16}}+4e^{\frac\beta\gamma\frac{3}{16}}&\to~~ 2^3 \nn\\
&Z_{4\times1}= e^{\frac\beta\gamma\frac{9}{4}}+3e^{\frac\beta\gamma\frac{7}{4}}+4e^{\frac\beta\gamma\frac{5}{4}}+3e^{\frac\beta\gamma\frac{3}{4}}+5e^{\frac\beta\gamma\frac{1}{4}}&\to~~ 2^4 \nn\\
&Z_{5\times1}= 2e^{\frac\beta\gamma\frac{53}{16}}+2e^{\frac\beta\gamma\frac{45}{16}}+6e^{\frac\beta\gamma\frac{37}{16}}+6e^{\frac\beta\gamma\frac{29}{16}}+6e^{\frac\beta\gamma\frac{21}{16}}+4e^{\frac\beta\gamma\frac{13}{16}}+6e^{\frac\beta\gamma\frac{5}{16}} &\to ~~ 2^5 \nn\\
&Z_{6\times1}=  e^{\frac\beta\gamma\frac{39}{8}}+3e^{\frac\beta\gamma\frac{35}{8}}+4e^{\frac\beta\gamma\frac{31}{8}}+7e^{\frac\beta\gamma\frac{27}{8}}+9e^{\frac\beta\gamma\frac{23}{8}}+11e^{\frac\beta\gamma\frac{19}{8}}+9e^{\frac\beta\gamma\frac{15}{8}}+8e^{\frac\beta\gamma\frac{11}{8}}+5e^{\frac\beta\gamma\frac{7}{8}}+7e^{\frac\beta\gamma\frac{3}{8}} &\to \underbrace{2^6}_{\beta\to0}\nn\\
\label{zss}
\end{eqnarray}
 We remark on the following features: \\
(i) The ground state is non-degenerate or doubly degenerate depending on $N$ being even or odd.\\
(ii) The spectrum is equally spaced $\Delta E=\beta/2\gamma$.\\
(iii) The highest energy level value is $E=-(\beta/\gamma)\,N/16$.\\
(iv) The number of distinct energy levels is
$$\text{N even:}~~~ \left(\frac N2 \right)^2+1,~~~~~~~~~~~ \text{N odd:}~~~ \left(\frac {N-1}2 \right)^2+\frac {N-1}2 +1 $$
(v) The identity \cite{SchW}
$$ \binom{N}{j}_q q^{j^2-j N}= \binom{N}{j}_{\frac 1 q} $$
allows us to reexpress the partition  function \eqref{Z1T} as 
$$ {Z}_{ N\times 1}=q^{-N/8}\sum_{j=0}^N\binom{N}{j}_{\frac 1 q} $$

\subsection{Thermodynamic  entropy}

\begin{figure}[tbp]
\begin{center}
{{\includegraphics[width=2.3in]{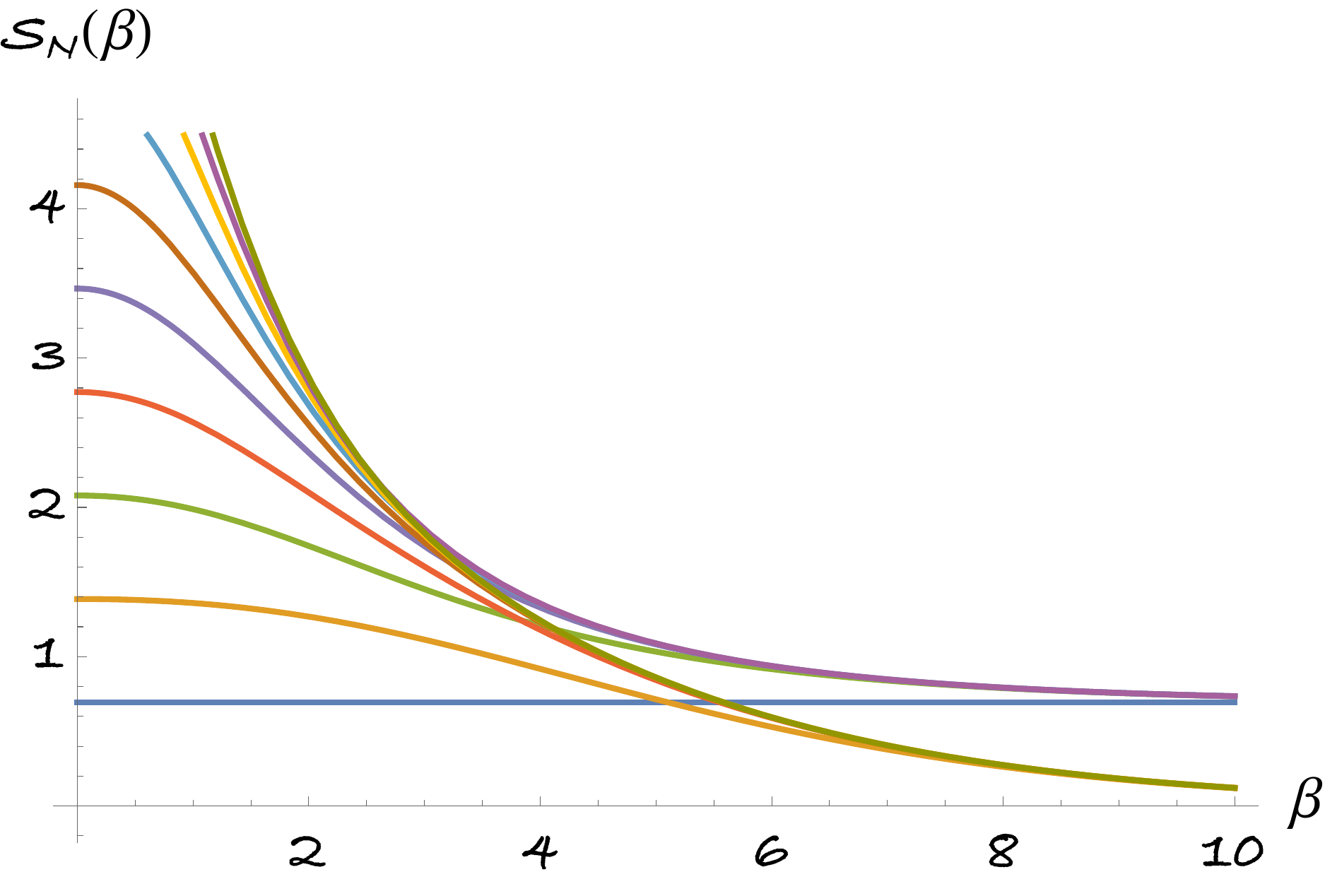} ~~~\includegraphics[width=3.1in]{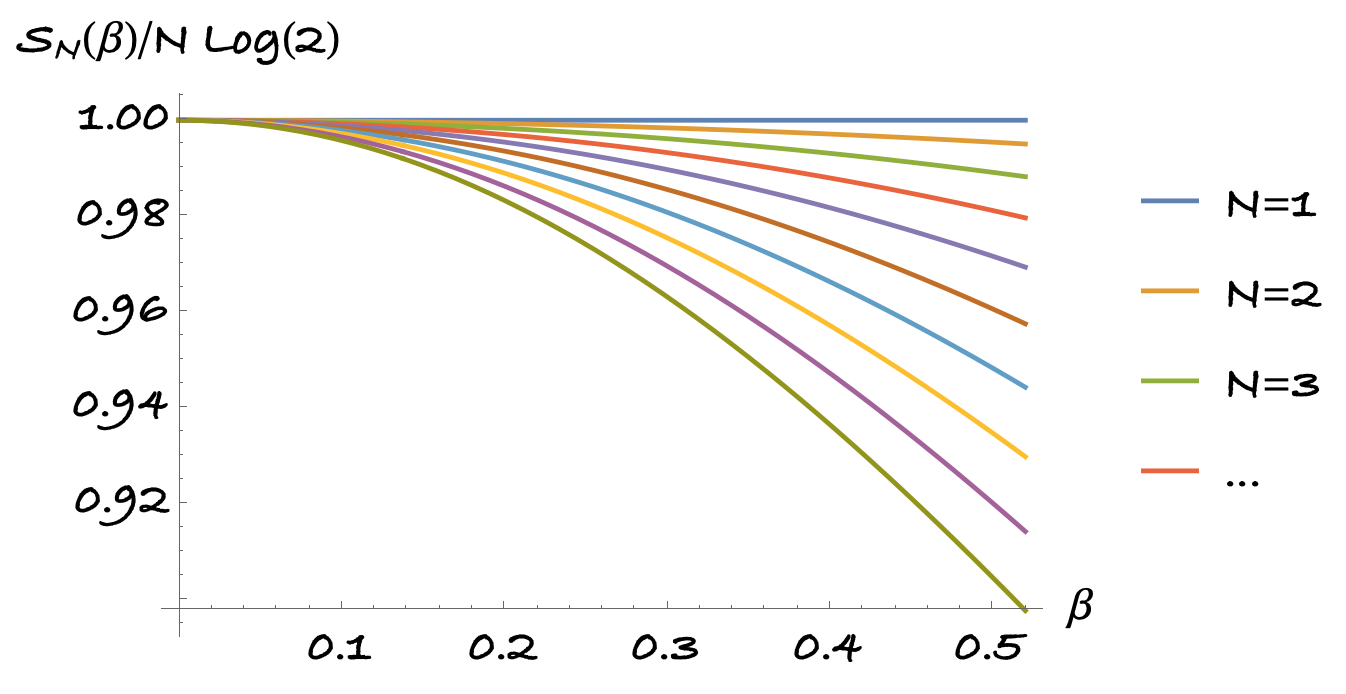}}}
\end{center}
\caption{{\sf Entropy as a function of temperature}. {\it Left}: at low temperature  the ground state is non-degenerate for odd $N$ ($S\to0$) and doubly degenerate  for even  $N$ ($S\to\log 2$). {\it Right}: in the high temperature limit  entropy asymptotes $S(0)=\log(2^N)$. Plots in the right show a positive specific heat $\partial S/\partial T>0$ as expected.} 
\label{entr}
\end{figure}

The partition function \eqref{Z1T} allows us to compute the entropy of the system from
$$
S(\beta)=(1-\beta\partial_\beta)\log Z(\beta)
$$  
 The entropy plots in fig.\ref{entr} show either a single or doubly degenerate ground state  at low temperature and consistency with the Hilbert space dimension $2^N$ at high energy.

\section{$L=1$ and Frahm+Polychronakos model }

\label{FPs}

\vspace{1mm}

\nin Modulo a normalization, which amounts to a zero point energy, the partition function \eqref{Z1T} coincides with that of the $SU(2)$ Frahm+Polychronakos AF model $Z_{FP}$. The relation  is
\be
 {Z}_{ N\times 1}=\underbrace{q^{-N^2/4-N/8}}_{\text{\sf zero point energy}}Z _{FP}
 \label{relation}
 \ee
where \cite{F93}, \cite{P93}
$$Z _{FP}=\sum_{j=0}^N\binom{N}{j}_q q^{(j-N/2)^2},~~~~~~~~~~ ~~~~q=e^{-\frac\beta{2\gamma}}  $$
The FP model is build from fundamental $SU(2)$ spins on a $N$-site chain with Calogero-type interactions. The  Hamiltonian can be written as \cite{F93}
\be
H_{FP} =\frac1{2\gamma}\left[- \sum_{k'>k} n_k n_{k'}+\sum_{k}  (1-k) n_k+\frac{N^2}{4}\right],~~~\qquad~~~k,k'=1,...,N
\label{PFHam}
\ee
with $k,k'$ denoting the lattice site and the occupation numbers $n_k= 0,1$. 

For later reference we quote the specific free energy of the FP model in the large $N$-limit. Defined as $Z_{FP}=e^{-\beta N {  f}_{FP}}$,  it reads \cite{F93}
\be
-\beta\, f_{FP} =T_{eff}\left(\frac{\pi^2}6+2\,f(1+e^{-\frac{\beta_{eff}}2})\right)~~~~~~~~\text{where}~~~~~f(x)=\int_1^xdy\frac{\log y}{1-y}.
\label{FF}
 \ee
where the dimensionless quantity    $\beta_{eff}\equiv\beta/2 \gamma'$. As mentioned at the beginning of section \ref{bs}, and pointed out by Frahm, a sensible large $N$-limit requires a $1/N$ scaling in the coupling $J$. Hence, \eqref{FF} arises from the replacement $\gamma= N\gamma'$ in \eqref{PFHam}.

\subsection{Fock space Hamiltonian: Casimirs and $SU(N)$ Cartans}

 We now find  the Fock space Hamiltonian which gives rise to the structure of energy levels of the $L=1$ model. It involves  the $SU(N)$ Casimir  as well as some definite chemical potentials for the $SU(N)$  Cartan charges.

 Mapping   Frahm's occupation numbers $n_k$ to the fermion flavor number, i.e. $n_k=\bar\Psi_k\Psi^k$ (no sum),  the $U(1)$ charge \eqref{u1} in Frahm's variables takes the form  ${\sf Q}=\sum_k n_k $. Then,
\begin{align}
 {\sf Q}^2 -N{\sf Q}  &= \sum_{k, k'}n_kn_{k'}-N\sum_k n_k\nn\\
&= \underbrace{n_1^2+n_2^2+...}_{=\sum_kn_k}+2n_1n_2+2n_1n_3+...-N\sum_k n_k\nn\\
&=2n_1n_2+2n_1n_3+...-(N-1)\sum_k n_k=2\sum_{k'>k}n_kn_{k'}-(N-1){\sf Q}
\end{align}
in the second line we used $n_k^2=n_k$ since $n_k=0,1$. From the relation \eqref{SUNU1} we conclude
\begin{align}
-\sum_{k'>k}n_kn_{k'}&=\frac{{\sf Q}-{\sf Q}^2}2\nn\\
&=\frac{N}{N+1}C^{(2)}+\frac{1-N}2{\sf Q}
\end{align}
This allows us to write the first term in \eqref{PFHam} in terms of the  $U(1)$ charge and  $SU(N)$   Casimir. 

The  term   $\sum_{k=1}^N k n_k$ in \eqref{PFHam} can be  be decomposed in Cartans $H_m$  and $\sf Q$ as (see App. \ref{Un} for conventions)
\begin{align}
&\left(
\begin{array}{ccccccc}
1&0&&\\
0&2&\\
&&3\\
&&&\ddots\\
&&&&N&\\
\end{array}
\right)=\mu^1{\mathbb I}+\mu^mH_m,~~~~~\qquad\qquad~~{m=2,...,N}. \nn
\end{align}
The solution for  $\mu$-coefficients is
$$\mu^1=\frac{N+1}2,~~~~\mu^m=-\sqrt{\frac{m(m-1)}2 }$$
The Fock space Hamiltonian in terms of the $SU(N)$ Casimir and conserved charges ${\sf Q}$ and  $Q_m\equiv\bar\Psi H_m\Psi=\bar\Psi_i(H_m)^i{}_{j}\Psi^j$ reads
\begin{align}
H_{FP} &=\frac{1}{2\gamma} \left[\frac N{N+1} C^{(2)}+\left( 1 - {N} \right){\sf Q} -  \mu^m Q_m + \frac{N^2}4 \right]\nn\\
&=\frac{1}{2\gamma} \left[-\frac12{\sf Q}^2+\left( 1 -\frac{N}{2}\right){\sf Q} -   \mu^m Q_m  + \frac{N^2}4 \right] .
\label{Cart}
\end{align} 
These expressions explicitly shows that non-singlet bosonization of the vector model, contrary to  singlet bosonization \eqref{bsn}, breaks $U(N)\to U(1)^N$.

\section{Large $N$} 
\label{LNs}
  
\subsection{$L=1$ (vector) model at large $N$: eigenvalue distribution}
 
It is clear from the eigenvalues integrals \eqref{partition} or \eqref{tie} that a sensible large $N$ limit, with potential and Vandermonde factor contributing at the same rate, can be obtained if we rescale $\gamma= N\gamma'$.  This rescaling agrees with that for $J$ in  \eqref{Hv}  and also with  the rescaling of energies done by Frahm in \cite{F93} and mentioned below \eqref{FF}.

To perform the large $N$ limit, we work in the representation given by \eqref{partition}. Our starting point is
\be
Z_{N\times 1} = \frac{2^N}{\int \prod_{i=1}^N d\lambda_i \prod_{j>i} \sinh^2 \left(\frac{\lambda_i-\lambda_j} 2  \right)   e^{-N\frac{ \gamma'}\beta  \lambda_i^2}} \int \prod_{i=1}^N d\lambda_i \prod_{j>i} \sinh^2 \left(\frac{\lambda_i-\lambda_j} 2  \right)  \cosh \left(\frac{\lambda_i} 2 \right) e^{-N\frac{\gamma'}\beta \lambda_i^2} 
\label{Vnn}
\ee
The $\cosh$ in this expression  scales as $\sim  O(N)$ and is therefore sub-leading wrt the $\sim O(N^2)$  potential and Vandermonde contributions.  Disregarding the $\cosh$, the matrix model  \eqref{Vnn}  corresponds to that of  $SU(N)$ Chern+Simons theory on $S^3$ \cite{M04}
\be
Z_{CS}^{\sf 3-sph} \propto  \int \prod_{i=1}^N d\lambda_i \prod_{j>i} \sinh^2 \left(\frac{\lambda_i-\lambda_j} 2  \right)   e^{- \frac N{2t} \lambda_i^2} .
\label{CS}
\ee
The CS eigenvalue distribution in the large $N$ limit is known to be given by \cite{A02},\cite{HY},\cite{M04}
\be
\rho_{CS}(\lambda)=\frac1{\pi t}{\rm arctan}\frac{\sqrt{e^t-(\cosh \lambda/2)^2}}{\cosh \lambda/2}.
\label{distr}
\ee
This is  a single-cut distribution supported on $\lambda\in[-2\,{\rm arcosh}\,e^{t/2},2\,{\rm arcosh}\,e^{t/2}]$. 

Hence, at large $N$,     the expectation value of $ \det(\cosh\frac M2)$ on the CS-matrix model can be computed as
\begin{align}
Z_{N\times 1}=2^N\langle\det\cosh\frac M2\rangle  \approx 2^N e^{ N \int_{\sf supp} d\lambda\, \log (\cosh(\lambda/2)) \rho_{CS}(\lambda)},~~~~~~~~N\to\infty
\label{LN}
\end{align}
where $t=\beta/2\gamma'=\beta_{eff} $.  From this last equation  we read the `planar' specific free  energy,  $f^{(0)}_{N\times1}\equiv F^{(0)}_{N\times1}/N$,
\be
-\beta \, f^{(0)}_{N\times1}=\ln2+2\times\frac{ T_{eff}}\pi\int_0^A\ln\big(\cosh\frac \lambda2\,\big)\:{\rm arctan}\frac{\sqrt{e^{\beta_{eff}}-(\cosh \lambda/2)^2}}{\cosh \lambda/2}\,d\lambda
\label{eigen}
\ee
where $A=2\,{\rm arcosh}\,e^{\beta_{eff}/2}$.  We have verified numerically the agreement  
$$\beta\, f_{N\times 1}=\beta\, f_{FP}-\frac{\beta_{eff}}4$$
  as expected from  \eqref{relation}.

\subsection{Large $N,L$ limit: eigenvalue distribution and thermal structure}
\label{tpf}
 
  We now consider a large $N\times L$ rectangular Fermionic matrix in the limit $\alpha\equiv L/N=$ finite.
 
\begin{figure}[tbp]
\begin{center}
{{\includegraphics[scale=0.4]{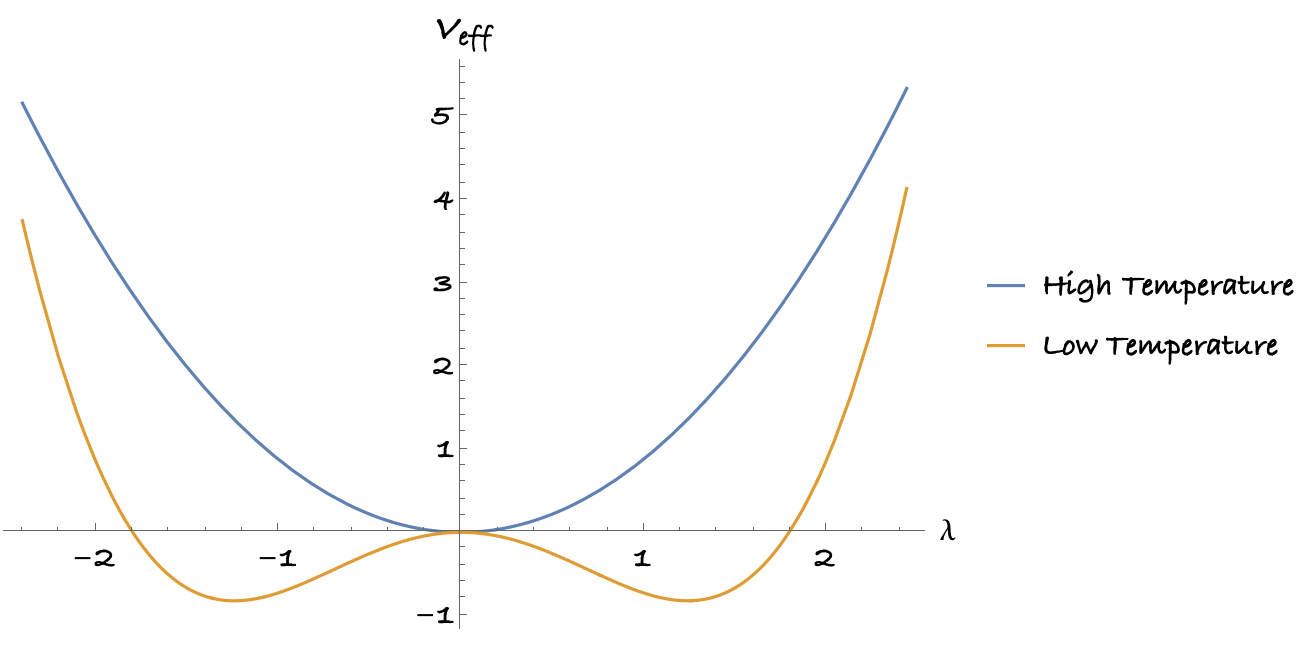} }}
\end{center}
\caption{ Effective eigenvalue potential \eqref{vefff} as a function of the temperature. The figure illustrates  symmetry breaking at low temperatures. As we show in the text, a single- to double-cut transition develops in the eigenvalue distribution.}
\label{thermo}
\end{figure}

The matrix integral \eqref{partition}, in the large $N,L\gg1$ limit,  is dominated by   saddle point   equations
\begin{equation}
\frac1N\sum_{j\ne i}\coth\left(\frac{\lambda_i-\lambda_j}2\right)=\alpha\left(2 \frac\gamma\beta\lambda_i-\frac
12\tanh\frac{\lambda_i}2\right),  
\label{discrete}
\end{equation}
the left-hand side arises from the Vandermonde repulsion and the right-hand side from the derivative of the effective potential 
\be
V_{eff}(\lambda)=\alpha\left(\frac\gamma\beta\lambda^2-\log\cosh\frac\lambda2\right).
\label{vefff}
\ee
Fig. \ref{thermo} displays its relevant features. In the high temperature regime ($ \gamma/\beta\gg1$) the quadratic piece localizes the eigenvalues near the origin,  the $\cosh$ becomes irrelevant and the Vandermonde can be accurately approximated by the standard one $\sinh^2\left( \frac{\lambda_i-\lambda_j}{2} \right)\sim\left( \frac{\lambda_i-\lambda_j}{2} \right)^2$. As temperature decreases a double well potential  develops  signaling  symmetry breaking. This phenomenon will show up in the eigenvalue distribution as  a transition from single  to double cut. Classically, the system becomes unstable for $\gamma/\beta<1/8$.  

Introducing the unit normalized eigenvalue distribution: 
\begin{equation}
\rho(\lambda)=\frac1N\sum_i\delta(\lambda-\lambda_i)\,,~~~~~~\int_{ C}\rho(\lambda)d\lambda=1
\label{nc}
\end{equation}
allows us to write  \eqref{discrete} in the form
\begin{equation}
\frac 1\alpha\,   - \hspace{-2mm}\!\!\!  \int_{ C} \coth\left( \frac{\lambda-\lambda'}2\right){\rho(\lambda')}\,d\lambda'= 2\frac\gamma\beta \lambda-\frac 12\tanh\frac{\lambda}2
,\quad \lambda \in {C}= {\sf supp}(\rho)\,.  
\label{inteq}
\end{equation}
Here $ - \hspace{-1.7mm}\!\!\!  \int$ denotes the  principal value prescription and $C$ denotes a curve in complex space. As evident from this expression, the spread of the  eigenvalues over the cut is measured by $\alpha$. As $\alpha\to\infty$, the repulsion term on the lhs becomes negligible and eigenvalues accumulate around the minimum of the potential. $\alpha$ is customarily called a t' Hooft coupling.

To solve the singular integral equation \eqref{inteq}  we change variables to $u=e^\lambda$ so that the lhs in \eqref{inteq} becomes the canonical Vandermonde repulsion. This modifies the normalization condition \eqref{nc} to 
\begin{equation}
\int_{C}\frac{du}u\tilde\rho(u)=1.
\label{normi}
\end{equation}
Here $\tilde\rho(u)\equiv\rho(\log u)$, and now $u\in(0,\infty)$. 
From the $\lambda \to-\lambda$ symmetry of the effective potential $V_{eff}$ we expect a symmetric distribution around the origin. Hence, in the high temperature regime, the single cut solution in $u$-variables  satisfies $u\in[a,b]$ with $ab=1$. Replacing $\coth\frac{\lambda-\lambda'}2=1+\frac{2u'}{u-u'}$ into \eqref{inteq} one finds
\begin{equation}
  -\hspace{-4mm}\int_C \frac{\tilde\rho(u')}{u'-u}\,du'= - \frac{2\alpha\gamma}\beta \log \sqrt u -\frac\alpha2\frac 1{u+1}+\delta,\quad
u\in C 
\label{int}
\end{equation}
with $ \delta= \alpha/4 +1/ 2$ a constant term. The $1/(u+1)$ term on the rhs arises from the $\cosh$ insertion in the  matrix integral \eqref{Vnn} (cf. (4.4) in \cite{A02}), setting it to zero reduces the problem to the Chern+Simons matrix integral.   

~

\subsection{Single cut   distribution}

\begin{figure}
\begin{center}
{{\includegraphics[width=3in]{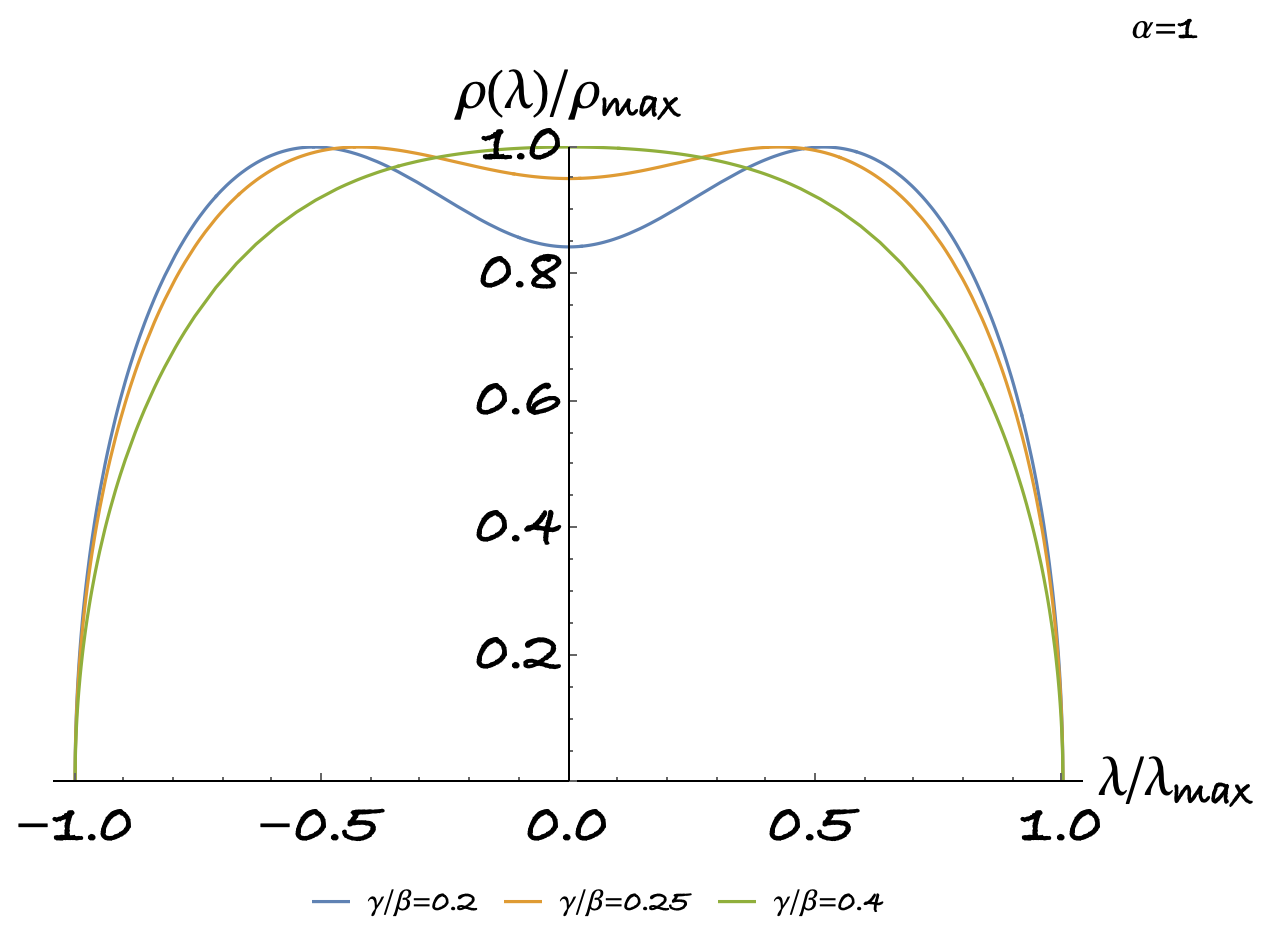} }}
\end{center}
\caption{{\sf 1-cut eigenvalue distribution as a function of Temperature}: the high temperature limit (green curve) of the eigenvalue distribution \eqref{rl}  resembles Wigner semi-circle law. As   temperature decreases the second term in \eqref{rl} becomes relevant and  a dip develops around the origin (blue curve). }
\label{lt}
\end{figure}
 
In the high temperature regime, i.e. $\gamma/\beta\gg1/8$, we expect a   single cut eigenvalue distribution  with support ${C}=[a,b],~~b>a>0$. 
Based on previous results \cite{A02},\cite{D15},  we propose\footnote{The ansatz \eqref{ansat} can be guessed if we deform the contour encircling the cut in  \eqref{mig} to  infinity. In this way we pick the pole at $\lambda$ and an integral over the discontinuity across the $\log$ branch cut, which suggests \eqref{ansat}.}
\begin{equation}
\tilde\rho(u) = \frac{1}{\pi} \left[  c_1 \tan^{-1}\left( \frac{ \sqrt{(u-a)(b-u)}}{u+\sqrt{ab}}\right) +c_2 \frac{\sqrt{(u-a)(b-u)}}{u+1} \right]~.
\label{ansat}
\end{equation}
Inserting this ansatz in \eqref{int} and using   \eqref{polos} and  \eqref{arctan} fixes $c_1 = 2\alpha\gamma/\beta$ and $c_2 = -\alpha/\left(2 \sqrt{(a+1)(b+1)}\right)$. The equations determining the endpoints $a$ and $b$ are those fixing the constant term $\delta$ in  \eqref{int} and the normalization condition for $\rho$. The former is
\begin{equation}  
\label{eqn1}
  \frac{2\alpha\gamma}\beta \log \left(\frac {\sqrt{a}+\sqrt{b}}2\right) +\frac\alpha{2\sqrt{(a+1)(b+1)}}= \frac\alpha4+\frac12
\end{equation}
while the latter, obtained inserting \eqref{ansat} in \eqref{normi} and using \eqref{c.1} and \eqref{n2}, results in 
\begin{equation}  
\label{eqn2}
\frac{2\alpha\gamma}\beta  \log \left(%
\frac{2 \sqrt{ab}+a+b}{4 \sqrt{ab}}\right) -\frac\alpha{2\sqrt{(a+1)(b+1)}}  \left(\sqrt{(a+1) (b+1)}-1-\sqrt{ab}\right) = 1~.
\end{equation}
It can be checked that \eqref{eqn1} and \eqref{eqn2} coincide when we set $a=1/b$. This relation is the expected relation for the endpoints  from the $\lambda_i \to - \lambda_i$ symmetry in \eqref{vefff}.  In summary, the single-cut eigenvalue distribution for the model \eqref{partition}  is
\be
\rho(\lambda) = \frac{\alpha}{\pi} \left[    \frac{2\gamma}\beta \tan^{-1}\left( \frac{ \sqrt{(e^\lambda-1/b)(b-e^\lambda)}}{e^\lambda+1}\right) -\frac{\sqrt b}{2(b+1)} \frac{\sqrt{(e^\lambda-1/b )(b-e^\lambda)}}{e^\lambda+1} \right]
\label{rl}
\ee
with $b$ satisfying
\begin{equation}
  \frac{2 \gamma}\beta \log \left(\frac{(b+1)^2}{4b} \right)- \frac{ \left(1-\sqrt b\right)^2}{2 (b+1)}=\frac1\alpha~.
  \label{a}
\end{equation}
Numerical methods are very effective in solving \eqref{a}.

\begin{figure}[tbp]
\begin{center}
\includegraphics[width=3in]{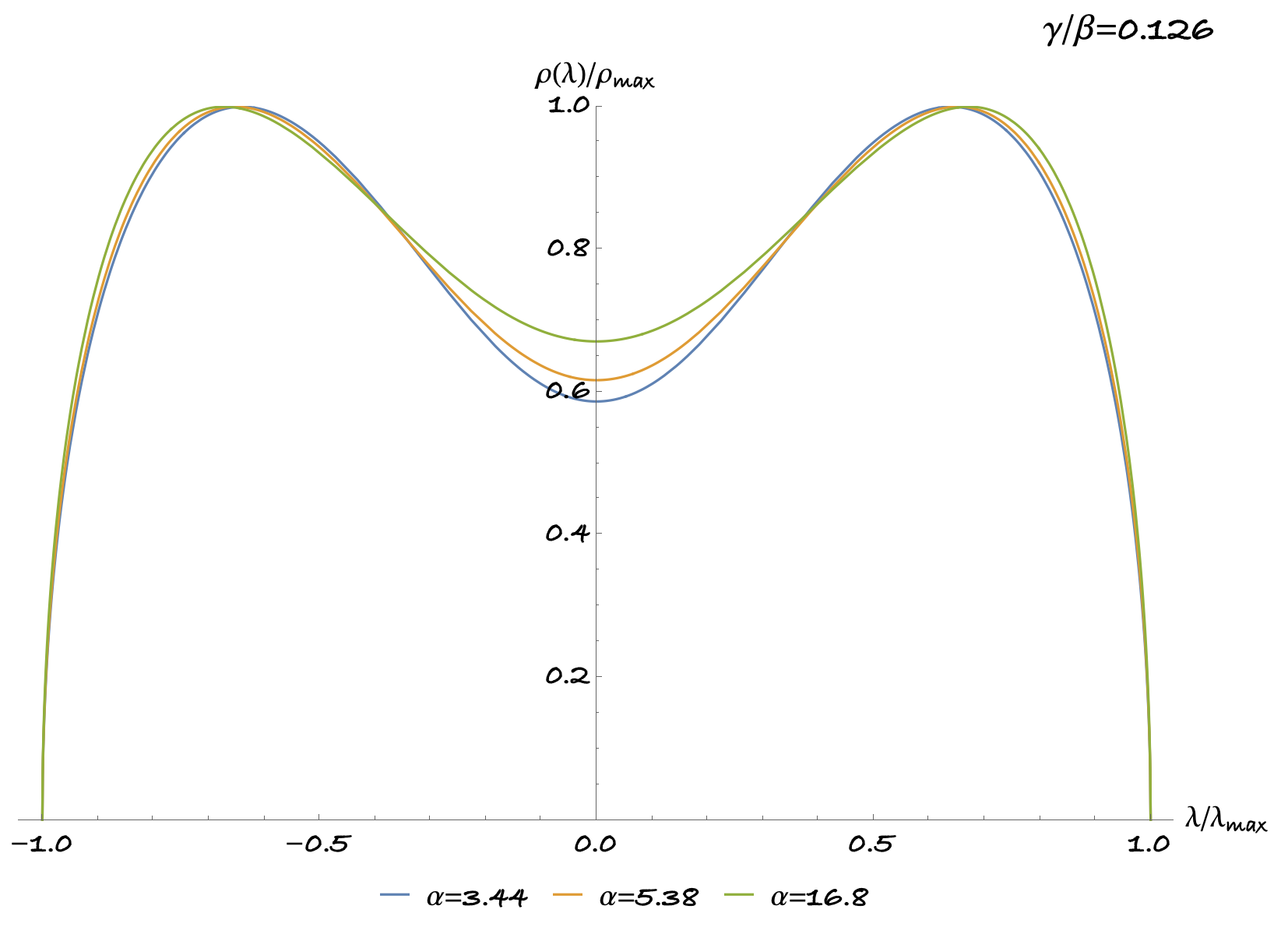} 
\includegraphics[width=3in]{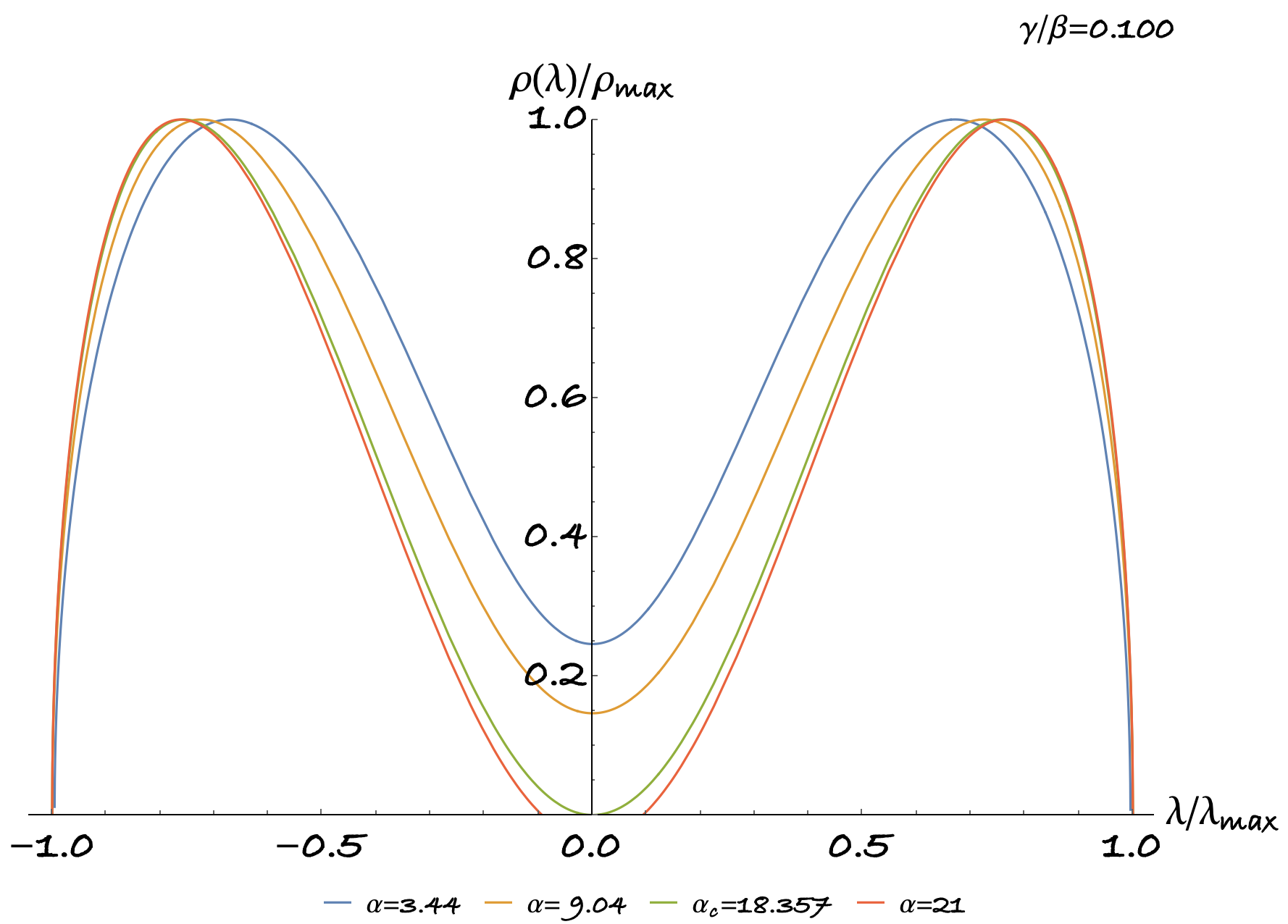} 
\end{center}
\par
\caption{ {\sf Left}: for   $\gamma/\beta>1/8$ the eigenvalue distribution is connected (single-cut) irrespectively of $\alpha$. {\sf Right}: for $\gamma/\beta<1/8$ and large enough $\alpha$ the distribution must be modified: as we increase $\alpha$ the dip reaches the origin and a disconnected (double-cut) eigenvalue distribution develops.}
\label{roLR}
\end{figure}

As we diminish the  repulsion, $\alpha\gg1$ (cf. \eqref{inteq}), eigenvalues locate near the potential minimum $\lambda=0$ ($ u=1$). In this regime the model effectively reduces to a quartic potential
\be
V_{eff}(\lambda)\approx \alpha\left(g\lambda^4+\mu \lambda^2\right),~~~~~~\alpha\gg1
\label{EF}
\ee
with $g=1/192$ and $\mu=\gamma/\beta-1/8$.  Then, the  single cut  endpoint can be found perturbatively in $1/\alpha$. Plugging  $b=1+\epsilon$ in \eqref{a} one finds
$$\epsilon=\frac{\beta/4\gamma}\alpha+O(\frac1{\alpha^2})$$

\subsection{Phase transition and double-cut solution}

\label{PTc}

We now discuss  some relevant features of the single-cut solution \eqref{rl}.   The two terms in \eqref{ansat} have support in the region $(1/b,b)$; however the first one is positive  while the second is negative. At high temperature   ($ \gamma/\beta\gg1$) the first term dominates and we  effectively get the Chern+Simons matrix integral with a Wigner semicircle-like  connected distribution (see fig. \ref{lt}). As we lower the temperature,  the second  term  becomes increasingly important  creating a dip in the semicircle (figs. \ref{lt} and \ref{roLR}).  For small enough temperature, the dip at the origin touches the horizontal axis. As a negative eigenvalue distribution is unacceptable, this signals a split of the support, i.e. a  single-    to double-cut  transition (see fig. \ref{roLR} right). 

\begin{figure}[tbp]
\begin{center}
\includegraphics[width=4in]{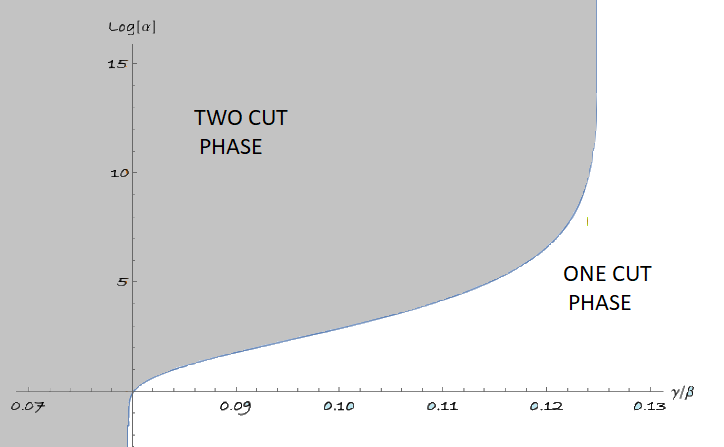}
\end{center}
\caption{{\sf Phase diagram}. Critical line showing single- to double-cut transition in the eigenvalue distribution:  to the left of the blue curve (low $T$) we have a disconnected (double-cut) distribution, to the right (high $T$) a connected (single-cut) one. The right asymptote lies at the classical value for instability $\gamma/\beta=1/8=.125$, the left asymptote at $\gamma/\beta=1/4\pi\approx 0.0795775$. The critical temperature for a square fermion matrix, $\alpha=1$, can be read from the critical line crossing the horizontal axis, this happens at $\gamma/\beta\approx0.0800517$. }
\label{trans}
\end{figure}

For fixed $\alpha$, the critical temperature $T_c$ below which the single- to double-cut transition occurs can be found by solving $\rho(0)=0$. This equation  gives a curve $\alpha(\gamma)$ in the $(\gamma,\alpha)-$plane which we display in fig. \ref{trans}. One concludes that for $\gamma/\beta>1/8$ the distribution is connected irrespectively of the value of $\alpha$ and for $\gamma/\beta<1/4\pi$ the distribution is disconnected irrespectively of $\alpha$.

\subsection{Order of the phase transition}

The order of the phase transition can be analytically studied in the $\alpha\gg1$ limit, where the system reduces to an effective $\lambda^4$ model (see Appendix \ref{q} for a review of the quartic model). 

Considering  the  first derivative of the specific free energy in the $\alpha\gg1$,
$$\frac{\partial f_0}{\partial g}=\lim_{N\to\infty}\frac1N\langle\alpha{\sf Tr}  M^4  \rangle=\alpha\int_C d\lambda\,\rho(\lambda)\,\lambda^4 $$
This expression can be easily computed by expressing the integral in terms of a contour $\cal C$ surrounding the cut $C$
$$\int_C d\lambda\,\rho(\lambda)f(\lambda)=\frac12\oint_{\cal C}dz \rho(z) f(z)$$
Then,
\be 
\frac{\partial f_0}{\partial g}=\left\{
\begin{array}{cc}
\frac{4\left(9g+\alpha\mu^2+ \mu\sqrt{\alpha^2\mu^2+12g\alpha}\right) }{\left(\sqrt\alpha\mu+\sqrt{\alpha\mu^2+12g}\right)^4 }    &~~ \mu>\mu_c\text{~~~1-cut}\\
\\
\frac{ \alpha\mu^2+g }{4g^2}&~~ \mu<\mu_c\text{~~~2-cut}
\end{array}
\right.
\ee
From this expression one finds evaluating at the critical value $\mu_c=-2\sqrt{g/\alpha}$ from both sides
$$\left.\frac{\partial f_0}{\partial g}\right|_{\mu=\mu_c^+}=\left.\frac{\partial f_0}{\partial g}\right|_{\mu=\mu_c^-}=\frac5{4g}, $$
$$\left.\frac{\partial^2 f_0}{\partial g^2}\right|_{\mu=\mu_c^+}=\left.\frac{\partial^2 f_0}{\partial g^2}\right|_{\mu=\mu_c^-}=-\frac9{4g^2}, $$ 
$$\left.\frac{\partial^3 f_0}{\partial g^3}\right|_{\mu=\mu_c^+}=\frac{27}{4g^3},~~~~~\left.\frac{\partial^3 f_0}{\partial g^3}\right|_{\mu=\mu_c^-}=\frac{13}{2g^3}, $$ 
Alternatively one can compute $\mu$-derivatives 
$$\frac{\partial f_0}{\partial \mu} =\alpha\int_C d\lambda\,\rho(\lambda)\,\lambda^2.$$  The results are
\be 
\frac{\partial f_0}{\partial \mu}=\left\{
\begin{array}{cc}
\frac{2\sqrt\alpha\left(8g+\alpha\mu^2+ \mu\sqrt{\alpha^2\mu^2+12g\alpha}\right) }{\left(\sqrt\alpha\mu+\sqrt{\alpha\mu^2+12g}\right)^3 }    &~~ \mu>\mu_c\text{~~~1-cut}\\
\\
-\frac{ \alpha\mu  }{2g}&~~ \mu<\mu_c\text{~~~2-cut}
\end{array},
\right.
\ee
$$\left.\frac{\partial f_0}{\partial \mu}\right|_{\mu=\mu_c^+}=\left.\frac{\partial f_0}{\partial \mu}\right|_{\mu=\mu_c^-}=\sqrt{\frac\alpha g}, $$
$$\left.\frac{\partial^2 f_0}{\partial \mu^2}\right|_{\mu=\mu_c^+}=\left.\frac{\partial^2 f_0}{\partial \mu^2}\right|_{\mu=\mu_c^-}=-\frac\alpha{2g }, $$ 
$$\left.\frac{\partial^3 f_0}{\partial \mu^3}\right|_{\mu=\mu_c^+}=\frac{\alpha^{3/2}}{4g^{3/2}},~~~~~\left.\frac{\partial^3 f_0}{\partial \mu^3}\right|_{\mu=\mu_c^-}=0. $$ 
The last result shows  a 3rd order phase transition.

Alternatively, one obtains the same behavior by  computing $\alpha$-derivatives of the specific free energy
$$\frac{\partial f_0}{\partial \alpha} =\int_C d\lambda\,\rho(\lambda)\,(g \lambda^4+\mu \lambda^2 ). $$ 
The results are
\be 
\frac{\partial f_0}{\partial \alpha}=\left\{
\begin{array}{cc}
\frac{1}{64} \alpha  \left(9 b^8 g^2+20 b^6 g \mu +8 b^4 \mu ^2\right)     &~~ \mu>\mu_c\text{~~~1-cut}\\
\\
\frac{1}{64} \alpha  g (a-b)^2 (a+b)^2 \left(8 \mu  \left(a^2+b^2\right)+g \left(5 a^4+6 a^2 b^2+5 b^4\right)\right)&~~ \mu<\mu_c\text{~~~2-cut}
\end{array}
\right.
\ee
$$\left.\frac{\partial f_0}{\partial \alpha}\right|_{\mu=\mu_c^+}=\left.\frac{\partial f_0}{\partial \alpha}\right|_{\mu=\mu_c^-}=-\frac{3}{4 \alpha }, $$
$$\left.\frac{\partial^2 f_0}{\partial \alpha^2}\right|_{\mu=\mu_c^+}=\left.\frac{\partial^2 f_0}{\partial \alpha^2}\right|_{\mu=\mu_c^-}=-\frac{1}{4 \alpha ^2}, $$ 
$$\left.\frac{\partial^3 f_0}{\partial \alpha^3}\right|_{\mu=\mu_c^+}=\frac{1}{4 \alpha ^3},~~~~~\left.\frac{\partial^3 f_0}{\partial \alpha^3}\right|_{\mu=\mu_c^-}=\frac{1}{2 \alpha ^3}. $$ 
\begin{figure}[tbp] 
\begin{centering}
{{\includegraphics[scale=0.6]{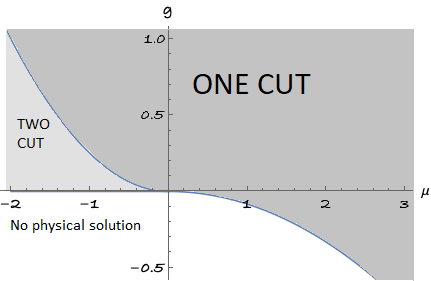} }}

\end{centering}
\caption{Phase diagram in $(\mu,g)$ displaying the 1- and 2-cut solutions.}
\label{Free}
\end{figure}
These results agree with the expected  third order behavior  originally recognized in \cite{GWW}. We display the phase diagram of the system in $(\mu,g)$-space in fig. \ref{Free}. 

\section {Summary }

In this paper we have studied fermionic  quantum mechanical models with  quartic interactions at finite temperature. The fermions have matrix character transforming as bi-fundamentals of $SU(N)$ and  $SU(L)$. The models can be solved exactly and several   large $N,L$ limits were performed. 

We have emphasized the role ambiguities play when performing the  bosonization technique. In particular we have shown that our regularization of the fermionic determinant in the non-singlet bosonization case breaks the classical symmetry from $U(N)\to U(1)^N$. In passing we have written Fermionic Fock space Hamiltonians  \eqref{H} and \eqref{Cart} showing that the spectrums found by bosonization in fact describe the precise Fermionic system at the quantum level with particular ambiguities being fixed.

For  finite $N,L$ case, the fermionic models were given an alternative representation in terms of vev's of characteristic polynomials in the Chern+Simmons matrix model \cite{T17}. Using  well-known orthogonal polynomials techniques (Berezin+Hikami formula \cite{BH}) the model can be solved solved. The vector model in non-singlet bosonization shows an equidistant spectrum and can be related to the (integrable) Polychronakos+Frahm model.   It is important to highlight that the bosonic matrix models we end up with, incorporate the temperature in a natural way  providing  an alternative viewpoint on the Chern+Simons matrix model. The results have also shown that the intricacies of the bosonization fix ambiguous normal terms to specific values in the Cartan generators (cf. \eqref{Cart}).

The eigenvalue distribution for the models in large $N,L$ limits were solved. For the vector model case ($L=1$) the distribution  coincides with the CS one and no phase transition arises as we vary temperature. Naturally, coupling constant   had to be rescaled for a sensible limit $\gamma\to N \gamma$. Fermionic matrices provide a new parameter (column/row ratio) which we can play with, i.e. $\alpha=L/N$ finite for  large $N,L$. The  eigenvalue distribution  in the large temperature (1-cut) regime was found in closed form (see \eqref{rl}). Its analysis shows, as expected, the splitting of the  cut as we lower the temperature due to the competition of two terms. For large $\alpha\gg1$ the model can be safely approximated by a quartic potential and well known results permit to find the 2-cut solution. One can check that a third order phase transition transition arises.  We also found the critical curve in $(\alpha,\gamma)$-parameter space where the  phase transition occurs.   

There are several avenues to explore in the model. The bosonic representation of the FP-model is novel to the authors knowledge and we find this an interesting avenue to explore. We can envisage studying the double scaling limit along the $\gamma(\alpha)$  curve to uncover the behavior of the bosonization technique. A more ambitious project is to try to study if the present fermionic models with finite Hilbert spaces can be related to de Sitter gravity or string models as elaborated in \cite{AM20},\cite{B21},\cite{S21},\cite{E21},\cite{AM21}. Finally we could study the $L=1$ model in the ferromagnetic case and see whether the transitions found in \cite{R20} take place in the present models.

~
 
\nin {\large\bf Ackwnoledgements} 
 
 \vspace{2mm}
 
\nin We thank Dio Anninos and Miguel Tierz for collaboration at the initial stages of this work.  We would like to thank  Beatrix M\"uhlmann and Miguel Tierz for many useful discussions and  we are very grateful to Dio Anninos and Jorge Russo for encouragement and support.  This work was funded by CONICET   grants PIP-1109-2017, PIP-UE 084  and UNLP grant X791. MS is supported by a CONICET fellowship. We would like to thank the referee for several useful comments.

\appendix

\section{$U(N)$ and $SU(N)$ conventions}   
\label{Un}

. $U(N)$: set of $N\times N$ complex matrices $U$ satisfying $UU^\dagger=1$.\\
. $SU(N)$: subset of $U(N)$ satisfying $\det U=1$.\\
. {\sf Fundamental representation}:   writing   
$U=e^{iA}$,  generators  $A$ are $N\times N$ hermitic matrices $A=A^\dagger$. \\
$ ~$~~ For $SU(N)$, the condition $~\det U=1\leadsto~$ traceless generators: $ {\sf Tr}A=0$. 
$$
SU(N)\text{ generators}:~~ \text{ traceless+hermitic ~ } \{\bm T_\alpha\} ~~~ ~~~~~\alpha=1,...,N^2-1
$$
Appending the identity to this set we obtain
$$
U(N)\text{ generators}:~~  \text{ hermitic ~ }\{ {\bm T}_{ a} \}=\left\{ \bm T_\alpha,\frac1{\sqrt{2N}}\mathbb I_N\right\},~~~~~~~~~~{ a}=1,...,N^2\nn 
$$
. {\sf Normalization}: fundamental generators are normalized as
\be
{\sf Tr}\big(\bm T_a\bm T_b\big)=\frac12\delta_{ab}
\label{norma}
\ee
. {\sf Killing+Cartan metric}: 
 $$ g_{ab}\equiv \delta_{ab} $$
. {\sf Fierz identity}: fundamental irrep generators satisfy the following identity
\be
 U(N): ~~~~\delta^{ab}(  T_{ a})^i{}_{ j}(  T_{b})^k{}_{ l}=\frac12  \delta^i_{ l}\delta^k_{j },  ~~\quad~~i,j=1,...N 
 \label{uN}
 \ee
Moving the identity generator in the lhs  of this last expression to the rhs we obtain
\be
SU(N) : ~~~~ \delta^{\alpha\beta}(T_\alpha)^i{}_{ j}(T_\beta)^k {}_{ l}=\frac12\left( \delta^i_{ l}\,\delta^k_{j }-\frac1N\delta^i_{ j}\,\delta^k_{ l} \right)
\label{fi}
\ee

---

{\footnotesize \nin {\sf Proof of \eqref{fi}}: the unit matrix $\bm 1$ and set $\bm T_\alpha$ form a basis for the space of hermitian $N\times N$ matrices, $\forall \bm A$ hermitian
$$\bm A=a^0 \bm 1+a^\alpha \bm T_\alpha$$
Taking the trace on both sides we determine $a^0$   
$${\sf Tr}\big(\bm A\big)=a^0 N~~\leadsto~~~a^0=\frac1N{\sf Tr}(\bm A)$$
Multiplying by $\bm T_\beta$ and taking the trace we find $a^\alpha$
$${\sf Tr}\big(\bm A\bm T_\beta\big)=a^0 \underbrace{{\sf Tr}\big(\bm T_\beta\big)}_{=0}+a^\alpha\underbrace{{\sf Tr}\big(\bm T_\alpha \bm T_\beta\big)}_{= \frac12g_{\alpha\beta}}~~\leadsto~~~a_\alpha=2{\sf Tr}(\bm A \bm T_\alpha)~~~\text{where}~~~a_\alpha\equiv g_{\alpha\beta}a^\beta$$
We conclude $\forall \bm A$
$${\bm A}=\frac1N{\sf Tr}\big(\bm A \big)\,\bm 1+2\, {\sf Tr}\big(\bm A\bm T^\alpha\big)\,\bm T_\alpha$$
where $\bm T^\alpha=g^{\alpha\beta}\bm T_\beta$. In components
$$  A^i{}_{ j}=\frac1N A^l{}_{ l}\, \delta^i_{ j}+ 2\, A^l{}_{ k}  (2\,T^\alpha)^k{}_{ l}\,\bm (T_\alpha)^i{}_{ j}~~~\leadsto ~~~A^l{}_{ k}\left((T^\alpha)^k{}_{ l}\,\bm (T_\alpha)^i{}_{ j}+\frac1N\delta^i_{ j}\,\delta^k_{ l}-\delta^i_{ l}\,\delta^k_{j }\right)=0~~~\leadsto~~~eq.~\eqref{fi}$$
This has to hold for arbitrary $\bm A$, therefore the Fierz identities \eqref{uN},\eqref{fi} follow.

---
}\\
. {\sf $SU(N)$ algebra}: structure constants are defined from fundamental (normalized) irrep generators as
$$[\bm T_\alpha,\bm T_\beta]=if^\gamma{}_{\alpha\beta}\bm T_\gamma$$ 
Multiplying by $\bm T^\rho$ and taking a trace we can express the structure constants in terms of traces of the fundamental irrep as
\be
 if^\rho{}_{\alpha\beta}=2\big[{\sf Tr}\big(\bm T^\rho\bm T_\alpha\bm T_\beta\big)-{\sf Tr}\big(\bm T^\rho\bm T_\beta\bm T_\alpha\big)\big]
\label{tr}
\ee

\nin . {\sf $SU(N)$  trace formulae}: from \eqref{fi} we obtain\\
\be
{\sf Tr}\left(\bm T^\alpha\bm X\right)\,{\sf Tr}\left(\bm T_\alpha\bm Y\right)=\frac12\left[{\sf Tr}\big(\bm X\,\bm Y\big)-\frac1N{\sf Tr}\big(\bm X \big){\sf Tr}\big(\bm Y \big)\right]
\label{tr1}
\ee
\be
{\sf Tr}\left(\bm T^\alpha\bm X\bm T_\alpha\bm Y\right)=\frac12\left[{\sf Tr}\big(\bm X \big){\sf Tr}\big(\bm Y\big)-\frac1N{\sf Tr}\big(\bm X  \bm Y \big)\right]
\label{tr2}
\ee 

\nin. {\sf $SU(N)$ quadratic Casimir for Fundamental and Adjoint irreps}:  the quadratic Casimir is defined as
\be
C_R^{(2)}\equiv g^{\alpha\beta}\,\bm T _\alpha(R)\,\bm T_ \beta(R)=\bm T^ \alpha(R)\bm T_ \alpha(R)
\label{casiC}
\ee
$~$ Fundamental irrep ${\tiny\yng(1)}$ :  \eqref{fi} gives
\be
\text{\sf Fundamental} :~~~~~~~~C^{(2)}_F=(T^\alpha T_\alpha)^i{}_{ j} =\frac{N^2-1}{2N}\delta^i_{j},~~~~~~~i,j=1,...,N
\label{FCas}
\ee
$~$ Adjoint irrep: generators are given by $(T_\alpha)^\rho{}_\sigma =if^\rho{}_{\alpha\sigma}$, then
\be
\text{\sf Adjoint}:~~~~~~~~~C^{(2)}_{\text{ adj}}=(T^\alpha T_\alpha)^\rho{}_{\sigma} =N\,\delta^\rho_{ \sigma}, ~~~~~~\alpha,\beta,.. =1,...,N^2-1
\label{AdI}
\ee

---

{\footnotesize \nin {\sf Proof}: 
\begin{align}
C^{(2)}_{\text{ adj}}&=\delta^{\alpha\beta}(if^\rho{}_{\alpha\gamma})(if^\gamma{}_{\beta\rho})\nn\\
&=4\big[{\sf Tr}\big(\bm T^\rho\bm T_\alpha\bm T_\gamma\big)-{\sf Tr}\big(\bm T^\rho\bm T_\gamma\bm T_\alpha \big)\big]\big[{\sf Tr}\big(\bm T^\gamma\bm T^\alpha \bm T_\sigma\big)-{\sf Tr}\big(\bm T^\gamma\bm T_\sigma \bm T^\alpha \big)\big]\nn\\
&=2\big[{\sf Tr}\big(\bm T_\gamma\bm T^\rho\bm T_\sigma \bm T^\gamma\big)
-{\sf Tr}\big(\bm T_\gamma \bm T^\rho\bm T^\gamma \bm T_\sigma \big)
-{\sf Tr}\big(\bm T^\rho\bm T_\gamma\bm T_\sigma \bm T^\gamma\big)+{\sf Tr}\big(\bm T^\rho\bm T_\gamma\bm T^\gamma \bm T_\sigma\big)\big]\nn\\
& =4\left[\frac{N^2-1}{4N }\delta^\rho_{\sigma}+\frac 1{4N} \delta^\rho_{\sigma} \right]=N\,\delta^\rho_{\sigma}\nn 
\end{align}
in going to the second line we used   \eqref{tr}, in going to the third  \eqref{tr1} and in the fourth we replaced $\bm T^\gamma\bm T_\gamma$ by $C^{(2)}_F$ and used \eqref{tr2}.

---
}\\
The young tableaux corresponding to the adjoint irrep involves $N$ boxes
$$\text{adj}=\left.\begin{array}{l}\tiny\yng(2,1)\\ \,\,\vdots\\ \tiny\yng(1)\end{array}\right\}N-1~\text{rows} $$
For completeness we quote the Casimir for a young tableaux $Y$ with $b$ boxes, row lengths $\rho_i$  and column lengths $\sigma_j$  (see app A in \cite{KT})
\be
C^{(2)}_Y=\frac12\left(bN+\sum_i\rho_i^2-\sum_j\sigma_j^2-\frac{b^2}N\right)
\label{genC}
\ee

%

\nin . {\sf Cartan generators}: for the fundamental representation of $SU(N)$ these are $N-1$ mutually commuting $N\times N$ traceless matrices $H_m ~ (m=2,...,N)$. We choose
$$\small\bm H_2=\alpha_2\left(
\begin{array}{ccccccc}
1&0&&\\
0&-1&\\
&&0\\
&&&\ddots\\
&&&&0\\
\end{array}
\right),~~ ~~
\bm H_3=\alpha_3\left(
\begin{array}{ccccccc}
1& 0&&\\
 0&1&0\\
&0&-2\\
&&&\ddots\\
&&&&0\\
\end{array}
\right)~,...,~~
\bm H_{N}=\alpha_N\left(
\begin{array}{ccccccc}
1& &&\\
 &1&\\
&&\ddots\\
&&&1&0\\
&&&0&-N+1\\
\end{array}
\right)$$
with $\alpha_m=1/\sqrt{2m(m-1)}$. They satisfy the normalization condition \eqref{norma}, i.e.  ${\sf Tr}(\bm H_m\bm H_n)=\frac12\delta_{mn}$.

\section{Brézin+Hikami vevs of characteristic polynomials}
\label{poly}
 
Expectation values of products of characteristic polynomials for Hermitian ensembles with probability distribution ${\cal P}(M)=\frac1{\cal Z}e^{-{\sf tr}\,V(M)}$ were computed in \cite{BH} in terms of orthogonal monic polynomials $m_n(x)$. These are orthogonal polynomials for $V(x)$ whose coefficients of the highest degree are equal
to unity 
$$\int m_n(x)m_m(x)\,e^{-V(x)}\,dx=h_n\delta_{nm},  
 \qquad\quad m_n(x)=x^n+\text{ lower degree}$$  The result is
\be
\langle\big(\det(\lambda-M)\big)^L\rangle_V=\frac{(-)^{L(L-1)/2}}{\prod_{i=0}^{L-1}(i!)}\det\left|
\begin{array}{cccc}
m_N(\lambda)&m_{N+1}(\lambda)&...&m_{N+L-1}(\lambda)\\
m'_N(\lambda)&m'_{N+1}(\lambda)&...&m'_{N+L-1}(\lambda)\\
\vdots\\
m^{(L-1)}_N(\lambda)&m^{(L-1)}_{N+1}(\lambda)&...&m^{(L-1)}_{N+L-1}(\lambda)
\end{array}\right|
\label{BHeqn}
\ee

\section{Some useful integrals}

We compute and quote here a number of integrals necessary for obtaining the eigenvalue distribution in section \ref{tpf}. They are obtained by standard applications of the residue theorem. Where  \eqref{int} is linear, the solution for $\rho$ can be found by superposition of the sources on the rhs. Results 1. and 2. solve the $1/(u+1)$ term, while results 3. and 4. solve the $\log{\sqrt u}$ term.

\vspace{2mm}

\nin 1.
\begin{equation}
\boxed{\frac{1}{\pi} \int_a^b \frac{dx}{x} \frac{\sqrt{(x-a)(b-x)}}{x+1} =
\left(\sqrt{(a+1) (b+1)} -\sqrt{ab}-1\right)~,\quad\quad 0<a<b.}  
\label{c.1}
\end{equation}
{\sf Proof}: the integrand in the lhs shows, in the complex plane, simple poles at $z=0,-1$ and a cut joining $z=A$ to $z=B$.  Let $\overline{\mathcal{C}}$ be a clockwise contour  surrounding the cut, then
\begin{equation}
\oint _{\overline{\mathcal{C}}} \frac{dz}z\frac{\sqrt{(z-a)(b-z)}}{z+1}%
=2i\int_a^b  \frac{dx}{x} \frac{\sqrt{(x-a)(b-x)}}{x+1}  
\label{c.2}
\end{equation}
Deforming the contour to infinity $\overline{\mathcal{C}}_{\infty}$, we pick two (residue) contributions from $z=-1$ and $z=0$. The result is
\begin{equation}
\oint_{\mathcal{C}_{-1}+\mathcal{C}_{0}+\overline{\mathcal{C}}_{\infty}} \frac{dz}z%
\frac{\sqrt{(z-a)(b-z)}}{z+1}=2\pi i\left[\sqrt{(a+1) (b+1)} -\sqrt{ab}-1%
\right]  \label{c3}
\end{equation}
Equating \eqref{c.2} and \eqref{c3} we obtain \eqref{c.1}.

\vspace{2mm}

\nin 2.
\begin{equation}
\boxed{\frac{1}{\pi} -\hspace{-3mm}\int_a^b \frac{dx}{x-y}
\frac{\sqrt{(x-a)(b-x)}}{1+x} = \frac{\sqrt{(a+1) (b+1)}}{y+1}-1~,\quad\quad
y\in (a,b)}  \label{polos}
\end{equation}
{\sf Proof}:  this integral is similar to \eqref{c.1}, but the pole at the origin is now located along the integration interval. Picking the same contour as before, surrounding  the cut in the clockwise sense, we have 
\begin{equation}
\oint_{\overline{\mathcal{C}}} \frac{dz}{z-y}\frac{\sqrt{(z-a)(b-z)}}{z+1}%
=2i\mathchoice {{\setbox0=\hbox{$\displaystyle{\textstyle -}{\int}$}
\vcenter{\hbox{$\textstyle -$}}\kern-.5\wd0}}
{{\setbox0=\hbox{$\textstyle{\scriptstyle -}{\int}$}
\vcenter{\hbox{$\scriptstyle -$}}\kern-.5\wd0}}
{{\setbox0=\hbox{$\scriptstyle{\scriptscriptstyle -}{\int}$}
\vcenter{\hbox{$\scriptscriptstyle -$}}\kern-.5\wd0}}
{{\setbox0=\hbox{$\scriptscriptstyle{\scriptscriptstyle -}{\int}$}
\vcenter{\hbox{$\scriptscriptstyle -$}}\kern-.5\wd0}} \!\int_a^b \frac{dx}{%
x-y} \frac{\sqrt{(x-a)(b-x)}}{1+x}  
\label{c4}
\end{equation}
Deforming the contour to infinity  we now pick a single contribution from $z=-1$, the final result is \eqref{polos}.

\nin 3.   Assuming $0<a<b $ are real numbers,
\begin{equation}
\boxed{ 
  \frac{1}{\pi} \int_a^{b} \frac{dx}{x} \tan^{-1} \frac{\sqrt{(x-a)(b-x)}}{x+\sqrt{ab}} = \log \left(\frac{2 \sqrt{ab}+a+b}{4 \sqrt{ab}}\right)~.}
\label{n2}
\end{equation}
4. Assuming $0<a<b $ are real numbers,
\begin{equation}
\boxed{\frac{1}{\pi} \,  -\hspace{-3mm} \int_a^{b} \frac{dx}{x-y} \tan^{-1} \frac{
\sqrt{(x-a)(b-x)}}{x+\sqrt{ab}} = - \log \left(\frac{2 \sqrt{y}}{\sqrt{a}+\sqrt{b}}\right)~.} 
\label{arctan}
\end{equation}
{\sf Proof}: 3. and 4. are generalizations of well known results of CS theory \cite{M04} after the change of variables $x=e^\lambda$. They can be checked with simple numerical examples.

\section{Quartic model \cite{M04,shi,molinari,bleher}}
\label{q}

The partition function for an hermitian matrix model,  writing  $M=U\cdot {\sf diag}(\lambda_1,...\lambda_N)\cdot U^{-1}$, is 
\begin{equation}
{\cal Z} = {\cal N} \int \prod_{i=1}^N d\lambda_i \prod_{j > i} \left( {\lambda_i-\lambda_j} \right)^2 e^{-N\,V(\lambda_i)} \,.
\label{Z4}
\end{equation}
 In the large $N$-limit the partition function is dominated by a saddle point equation  for the eigenvalue distribution $\rho(\lambda)$  (cf. \eqref{nc})   
\begin{equation}
   2 - \hspace{-2mm}\!\!\!  \int_{ C} \frac{\rho(\lambda')}{\lambda-\lambda'} d\lambda'=V'(\lambda)
,\quad \lambda \in {C}= {\sf supp}(\rho)\,.  
\label{inteq2}
\end{equation}
The solution to this equation is found by introducing the resolvent $\omega(z)\equiv \frac 1 N\langle{\sf Tr}\frac1{z-M}\rangle$ with $z\in\mathbb C$, which in the large $N$-limit takes the form
$$\omega_0(z)=    \int_{ C} \frac{\rho(\lambda')}{z-\lambda'} d\lambda'$$
Away from $C$ (support of the distribution) this is an analytic function.  The resolvent satisfies three important conditions  \cite{BIPZ,M04} : 

(i)  $\rho(\lambda)=-\frac1{2\pi i}(\omega_0(\lambda+i\epsilon)-\omega_0(\lambda-i\epsilon))$ for $\lambda\in{\sf supp} (\rho)$,

 (ii) $\omega_0(\lambda+i\epsilon)+\omega_0(\lambda-i\epsilon)=V'(\lambda)$ 
 
 (iii)  $\omega_0(\lambda)=\frac1\lambda+O(\frac1{\lambda^2})$ as $\lambda\to\infty$.\\
This turns the integral equation \eqref{inteq2} into a Riemann+Hilbert problem for $\omega(\lambda)$.

We are interested in analyzing the consequences of symmetry breaking on the eigenvalue distribution. To this end we consider the potential   
\be
V (\lambda)=\alpha\left(g\lambda^4+\mu \lambda^2\right)\,.
\label{4}
\ee
and  assume $g>0$. Classically the potential develops two minima at $\lambda_{min}=\pm\sqrt{-\frac\mu{2g}}$ when $\mu$ becomes negative. These break the $\mathbb Z_2$ symmetry  of the potential. 
In the large $N$-limit, the instability develops below $\mu_c=-2\sqrt {g/\alpha}<0$  when the distribution becomes disconnected (2-cut) \cite{shi,molinari,bleher}. 

Explicitly,  the quartic model requires to solve
$$\frac1\alpha    - \hspace{-2mm}\!\!\!  \int_{ C} \frac{\rho(\lambda')}{\lambda-\lambda'} d\lambda'=2 g\lambda^3+\mu\lambda,$$
Notice $\alpha$   measures the strength of the eigenvalue repulsion (lhs). As $\alpha\to\infty$ the repulsion is negligible and eigenvalues localize at the extrema of $V(\lambda)$. Hence, the size of the cuts is controlled by $\alpha$.

\subsection{1-cut solutions}

The 1-cut solution $\omega_0^{\sf 1-cut}(z)$  for arbitrary $V(\lambda)$  satisfying (i)-(iii) above can be written   in closed form as \cite{migdal} (see also \cite{Ake96})
\be
\omega_0^{\sf 1-cut}(z)=\frac1{2}\oint_{\cal C}\frac {dw}{2\pi i}\frac{V'(w)}{z-w}\left(\frac{(z-a)(z -b)}{(w-a)(w-b)}\right)^{1/2}
\label{mig}
\ee
here $\cal C$ is an anti-clockwise contour encircling the (single) cut $C=[a,b]$, $a<b$. 

The endpoints of the cut, $a$ and $b$,  become fixed upon demanding the resolvent $\omega$ to satisfy property (iii). Expanding the   integrand in \eqref{mig} for $z\gg w$, condition (iii) implies that 
\be
 \oint_{\cal C}\frac {dw}{2\pi i}\frac{V'(w)}  {\sqrt{ (w - b)(w-a)}}=0 ~~~~~\text{and}~~~~~\frac12\oint_{\cal C}\frac {dw}{2\pi i}\frac{(w-(a+b)/2)\,V'(w)}  {\sqrt{(w-a)( w - b) }}=1 .
\label{1c}
\ee
Inserting 
$$V'(w)=2\alpha\left(2 gw^3+ \mu w\right),$$ 
integrals \eqref{1c} can be easily calculated deforming the contour $\cal C$ to infinity. The circulation at infinity   picks the coefficient of  the $1/w$-term in the expansion of the integrand, i.e residue at infinity. The results are
\be
 (a+b)\left((5 a^2 -2 a  b +  5 b^2) g + 4   \mu\right)  =0
 \label{c1}
 \ee
\be
\frac\alpha{64} (a-b)^2\left((15 a^2  + 18 a b    + 15 b^2) g + 8  \mu \right)=1.
   \label{c2}
\ee 
A. {\sf Symmetric solution}:   \eqref{c1} is immediately solved for    $a=-b$, which inserted in \eqref{c2} gives 
\be
b^2=\frac4{\alpha\mu+\sqrt{\alpha^2\mu^2+12\alpha g} },
\label{b1c}
\ee
As expected, the size of the cut reduces, $b\to0$, when the eigenvalue repulsion diminishes $\alpha\to\infty$. 

Deforming the contour $\cal C$ in \eqref{mig} to infinity, $\oint_{\cal C}=\oint_{\bar{\cal C}_z}+\oint_{{\cal C}_\infty}$, one obtains
\be
\omega_{0}^{\sf 1sym}(z)= \alpha\left(2 gz^3+\mu z\right)-\alpha\sqrt{z^2-b^2 }\left(2gz^2+b^2g+\mu\right)\,.
\label{res}
\ee
The first term comes from the pole at $w=z$ and the second from the circulation at infinity. In the limit $g=0$ the solution \eqref{b1c}-\eqref{res} smoothly reduces to Wigner's semicircle.

 The second term in \eqref{res} is responsible for the discontinuity required by property (i) above; hence, we immediately read the single-cut eigenvalue distribution for the quartic model \cite{BIPZ,shi,molinari,bleher,M04} 
\be
\rho^{\sf 1sym}(\lambda)=\frac\alpha\pi\sqrt{b^2-\lambda^2 }\left(2g\lambda^2+b^2g+\mu\right) ~~~~~\text{for}~~~\lambda\in C=[-b,b]
\label{reg}
\ee
This solution is consistent as long as $\rho^{\sf 1s}\ge0$. For $g>0$, a dip develops around the origin as $\mu$ becomes negative  (cf. Fig. \ref{lt}). At $\mu_c=-2\sqrt {g/\alpha}$ the distribution touches the horizontal axis. For $\mu<\mu_c$ the eigenvalue distribution consists of two disconnected cuts,  we recompute it below.  \\
In summary, solution \eqref{b1c}-\eqref{res} makes sense for:\\
~~ . $\mu>0\leadsto g>-\frac{\alpha  \mu^2}{12} $. This condition follows from demanding   $b\in\mathbb R$ \cite{BIPZ}. Amusingly, although the potential becomes
unbounded below for  $g<0$, as long as $g>-\frac{\alpha  \mu^2}{12} $  the potential barrier   prevents the eigenvalues  from overflowing. At the critical point $g = -\alpha \mu^2/12$, the non-analytic behavior of $\rho^{\sf 1sym}(\lambda)$ at the edge of the support  changes from $|\lambda \pm b|^{1/2}$ to $|\lambda \pm b|^{3/2}$ . This 
phenomenon is crucial in 2d quantum gravity and non-critical string theory. \\
~~ . $\mu\le0,g>0\leadsto  \mu\ge\mu_c=-2\sqrt {g/\alpha}$ or equivalently $g\ge\frac{\alpha\mu^2}{ 4}$. The condition follows demanding that  $\rho^{\sf 1s} \ge0$.

\vspace{2mm}
\nin B. {\sf Asymmetric solution}: for $g>0$ and $\mu<0$, equations \eqref{c1}-\eqref{c2} allow for a single cut solution centered close to one of the minima $\lambda_{min}=\pm\sqrt{-\frac\mu{2g}}$. Inserting $a=\sigma-\tau$ and $b=\sigma+\tau$  in \eqref{c1}-\eqref{c2} the solution is \cite{shi}
\be
 \sigma^2 =\frac{-3\mu + 2  \sqrt{ \mu^2- \frac {15g} \alpha}}{10g},~~~~~~~\tau^2=\frac{-2\mu - 2  \sqrt{ \mu^2- \frac {15g} \alpha}}{15g}
 \label{st}
 \ee
As repulsion vanishes, $\alpha\to\infty$, the `center' of the distribution $\sigma\to\lambda_{min}$ and the width $\tau\to0$. The solution runs away $\sigma\to\infty$ as $g\to0$. Deforming the contour $\cal C$ in \eqref{mig} to infinity one obtains
\be
\omega_{0}^{\sf 1asym}(z)= \alpha\left(2 gz^3+\mu z\right)-\alpha\sqrt{(z -\sigma-\tau)(z-\sigma+\tau) }\left(2gz^2+2gz\sigma+2g\sigma^2+\tau^2+\mu\right)\,.
\label{res2}
\ee
The eigenvalue distribution reads
\be
\rho^{\sf 1asym}(\lambda)=\frac\alpha\pi\sqrt{(\sigma+\tau-\lambda)(\lambda-\sigma+\tau) }\left(2g\lambda^2+2g\lambda\sigma+2g\sigma^2+\tau^2+\mu\right)~~~~~\text{for}~~~\lambda\in C=[\sigma-\tau,\sigma+\tau ]
\label{rA}
\ee
In summary, the asymmetric solution \eqref{st}-\eqref{rA} exists for the following:\\
~~ . $\mu<0\leadsto 0<g<\frac{\alpha  \mu^2}{15} $. This condition arises from $\sigma,\tau\in\mathbb R$. \\
Notice the asymmetric solution \eqref{rA} gives rise to $\langle{\sf Tr} M\rangle\ne0$ as compared to \eqref{reg}. For further discussions see \cite{shi}.

\subsection{Two cut solutions}

For $g>0,\mu<0$ and  $\mu\le\mu_c $ the support of the distribution becomes disconnected.
\vspace{2mm}

\nin {\sf A. Symmetric solution}: multicut solutions can also be written in closed form \cite{M04}. For even potentials $V(\lambda)=V(-\lambda)$, the ($\mathbb Z_2 $ symmetric) 2-cut solution takes the form
\be
\omega_0^{\sf 2-cut}(z)=\frac1{2}\oint_{\cal C}\frac {dw}{2\pi i}\frac{V'(w)}{z-w}\left(\frac{(z^2-a^2)(z^2 -b^2)}{(w^2-a^2)(w^2-b^2)}\right)^{1/2}
\label{mig2}
\ee
Here, $\cal C$ is an anti-clockwise circulation around the cuts  located symmetrically around the origin at $-b<x <-a$ and $a<x<b$ with $b>a>0$.

Demanding the resolvent to satisfy property (iii) one obtains
$$ \oint_{\cal C}\frac {dw}{2\pi i}\frac{ V'(w)}  {\sqrt{ (w^2 - b^2)(w^2-a^2)}}=0 ~~~~~\text{and}~~~~~\frac12\oint_{\cal C}\frac {dw}{2\pi i}\frac{(w^2-(a^2+b^2)/2)\,V'(w)}  {\sqrt{(w^2-a^2)( w^2 - b^2) }}=1 $$
Inserting the explicit expression for $V'(w)$ and computing the contour integral at infinity one finds
\begin{align} 
(a^2 + b^2) g + \mu=0,~~~~~\frac{g\alpha}4(a^2 -    b^2 )^2=1 ~~\leadsto~~a^2=-\frac\mu{2g}-\frac1{\sqrt{\alpha g}},~~
b^2=-\frac\mu{2g}+\frac1{\sqrt{\alpha g}}
\end{align}
The 2-cuts $\mathbb Z_2$-symmetric solution shows two peaks centered around the classical minima $\lambda_{min}^2=-\frac\mu{2g}$ with widths scaling as $1/\alpha$. As we increase the eigenvalue repulsion $\alpha\to0$, the peaks widen  and become eventually connected for $\alpha<4g/\mu^2$.

To find the resolvent we deform $\cal C$ to infinity. From the pole at $w=z$ and the residue at infinity in \eqref{mig2} one obtains
$$\omega_0^{\sf 2s}(z)= \alpha\left(2 gz^3+\mu z\right)-2g\alpha\sqrt{(z^2-a^2)(z^2 -b^2)}z$$
The eigenvalues distribution follows from the second term
\be
\rho^{\sf 2s}(\lambda)=\frac{2g\alpha}\pi\sqrt{(b^2-\lambda^2)(\lambda^2-a^2) }|\lambda| ~~~~~\text{for}~~~\lambda\in C=[-b,-a]\cup[a,b]
\label{reg2}
\ee

\section{Frahm+Polychronakos model}

The Fram+Polychronakos model  arises from a particular limit of  a Calogero-type spin system \cite{calo}.  The FP model consists of fundamental $SU(n)$ spins positioned at the equilibrium positions of a classical Calogero model, i.e.  interacting through inverse–square exchange. In the present paper, we will be concerned with the $SU(2)$ case.

Start with an $N$-particles system on the line with the Hamiltonian \cite{P92i}
\begin{equation}
H=\frac12 \sum_i (p_i^2+\omega^2 x_i^2)+ \sum_{i<j} \frac{l(l-M_{ij})}{(x_i-x_j)^2}
\label{polcal}
\end{equation}
where $M_{ij}$ is the particle permutation operator 
$$M_{ij}=M_{ji}=M_{ij}^\dagger,~~~M_{ij}^2=1$$ $$M_{ij}A_j=A_iM_{ij},~~~M_{ij}A_k=A_kM_{ij},~~\text{for }k\ne i,j$$
where $A_i$ is any operator (including $M_{ij}$ themselves) carrying one or more particle indices. This model was shown to be integrable in \cite{P92i}. Rescaling  $\omega\to l\omega$,  the Hamiltonian naturally splits into  Calogero and  spin parts, $H=H_C+H_s$, with
\be
H_C=\frac12 \sum_i (p_i^2+l^2\omega^2 x_i^2)+ \sum_{i<j} \frac{l^2  }{(x_i-x_j)^2},~~~~~H_{s}=-l  \sum_{i<j} \frac{ M_{ij}}{(x_i-x_j)^2}
\label{Ll}
\ee
In the strong coupling limit $l\to\infty$, the coordinate degrees of freedom freeze and decouple from the spin ones. Setting the coordinates to their  static equilibrium positions
\be
\omega^2 x_i=2\sum_{j\ne i}\frac1{(x_i-x_j)^3}
\label{Hermi}
\ee
we end up with a spin chain of  fundamental $SU(n)$ spins  lying at the equilibrium positions of the classical Calogero system. The solutions $x_i$ of \eqref{Hermi} can be identified as  the $N$ roots of the Hermite polynomial $H_N(x)$ of degree $N$ \cite{sz}.

The Frahm+Polychronakos model relevant for this paper is defined as the fermionic spin chain arising from the spin part   in \eqref{Ll} with $M_{ij}\mapsto -\sigma_{ij}$. Dropping the $l$ factor, 
  $$H_{FP}\equiv \frac1{2\gamma}H_s$$
The partition function for the fermionic $SU(2)$ model was guessed numerically in \cite{F93} and derived analytically in \cite{P93}
\begin{equation}
Z_{2,N}^{af}=q^{-\frac{m}{4}} \sum_{k=0}^N q^{(k-\frac{N}{2})^2} \prod_{r=1}^k \frac{1-q^{N-r+1}}{1-q^r}~~~\text{with}~~~q=e^{-\frac\beta{2\gamma}} .
\label{z2n}
\end{equation}

   \eject

\bibliographystyle{apsrev4-1long}
\bibliography{GeneralBibliographyTwo.bib}
\end{document}